\newif\ifsubmode
\submodefalse

\newif\ifprintfig
\printfigfalse

\ifsubmode
  \documentstyle[12pt,aasms4,epsf]{article}
  \received{}
  \accepted{}
  \journalid{}{}
  \articleid{}{}
\else
  \documentstyle[11pt,aaspp4,epsf]{article}
  \slugcomment{{\sl submitted to The Astrophysical Journal; May 12, 1997.}}
\fi

\lefthead{van der Marel et al.}
\righthead{A black hole in M32 --- II.~axisymmetric dynamical models}

\def\etal{{et al.~}}
\def\lta{\lesssim}
\def\gta{\gtrsim}
\def\kms{\>{\rm km}\,{\rm s}^{-1}}
\def\ergs{\>{\rm erg}\,{\rm s}^{-1}}
\def\pc{\>{\rm pc}}
\def\yr{\>{\rm yr}}
\def\cm{\>{\rm cm}}
\def\Mpc{\>{\rm Mpc}}
\def\Msun{\>{\rm M_{\odot}}}
\def\Lsun{\>{\rm L_{\odot}}}

\def\Mdark{M_{\bullet}}
\def\Rc{R_{\rm c}}
\def\vrsqav{\langle v_r^2 \rangle}
\def\vthsqav{\langle v_{\theta}^2 \rangle}

\def\reffHSTBHpred{2}
\def\reffchisqBH{3}
\def\reffzvcplot{5}
\def\reffcontour{6}
\def\reffbestfits{7}
\def\reffdatafitsAtoH{8}
\def\reffmoments{10}

\begin{document}

\title{Improved evidence for a black hole in M32 from HST/FOS spectra ---\\
       II.~Axisymmetric dynamical models\altaffilmark{1}}

\author{Roeland P.~van der Marel\altaffilmark{2}}
\affil{Institute for Advanced Study, Olden Lane, Princeton, NJ 08540}

\author{N.~Cretton, P.~Tim de Zeeuw,}
\affil{Sterrewacht Leiden, Postbus 9513, 2300 RA Leiden, The Netherlands}

\author{Hans--Walter Rix\altaffilmark{3}}
\affil{Steward Observatory, University of Arizona, Tucson, 
       AZ 85721}


\altaffiltext{1}{Based on observations with the NASA/ESA Hubble Space
       Telescope obtained at the Space Telescope Science Institute,
       which is operated by the Association of Universities for
       Research in Astronomy, Incorporated, under NASA contract
       NAS5-26555.}

\altaffiltext{2}{Hubble Fellow.}

\altaffiltext{3}{Alfred P.~Sloan Fellow.}


\ifsubmode\else
\clearpage\fi


\ifsubmode\else
\baselineskip=14pt
\fi


\begin{abstract}
Axisymmetric dynamical models are constructed for the E3 galaxy M32 to
interpret high spatial resolution stellar kinematical data obtained
with the Hubble Space Telescope (HST). Models are studied with
two-integral, $f(E,L_z)$, phase-space distribution functions, and with
fully general three-integral distribution functions. The latter are
built using an extension of Schwarzschild's approach: individual
orbits in the axisymmetric potential are calculated numerically, and
populated using non-negative least-squares fitting so as to reproduce
all available kinematical data, including line-of-sight velocity
profile shapes. The details of this method are described in companion
papers by Rix et al.~and Cretton et al.

Models are constructed for inclinations $i=90^{\circ}$ (edge-on) and
$i=55^{\circ}$. No model without a nuclear dark object can fit the
combined ground-based and HST data, independent of the dynamical
structure of M32. Models with a nuclear dark object of mass $\Mdark =
3.4 \times 10^6 \Msun$ (with $1\sigma$ and $3\sigma$ error bars of
$0.7 \times 10^6 \Msun$ and $1.6 \times 10^6 \Msun$, respectively) do
provide an excellent fit. The inclined models provide the best fit,
but the inferred $\Mdark$ does not depend sensitively on the assumed
inclination. The models that best fit the data are not two-integral
models, but like two-integral models they are azimuthally
anisotropic. Two-integral models therefore provide useful low-order
approximations to the dynamical structure of M32. We use them to show
that an extended dark object can fit the data only if its half-mass
radius is $r_{\rm h} \lta 0.08''$ ($=0.26 \pc$), implying a central
dark matter density exceeding $1 \times 10^8 \Msun \pc^{-3}$.

The inferred $\Mdark$ is consistent with that suggested previously by
ground-based kinematical data. However, radially anisotropic
axisymmetric constant mass-to-light ratio models are now ruled out for
the first time, and the limit on the dark matter density implied by
the HST data is now stringent enough to rule out most plausible
alternatives to a massive black hole. Thus, the evidence for a massive
black hole in the quiescent galaxy M32 is now very compelling.

The dynamically inferred $\Mdark$ is identical to that suggested by
existing models for HST photometry of M32 that assume adiabatic
growth (over a time scale exceeding $10^6 \yr$) of a black hole into a
pre-existing core. The low activity of the nucleus of M32 implies
either that only a very small fraction of the gas that is shed by
evolving stars is accreted onto the black hole, or alternatively, that
accretion proceeds at very low efficiency, e.g.~in an
advection-dominated mode.
\end{abstract}



\clearpage


\section{Introduction}

It is generally believed that active galaxies and quasars are powered
by the presence of massive black holes (BHs) in their nuclei, and that
such BHs are present in many, possibly all, quiescent galaxies as well
(see Kormendy \& Richstone 1995, Lynden-Bell 1996 and Rees 1996 for
reviews of this paradigm and its history). Evidence for this can be
derived from studies of the dynamics of stars and gas in the nuclei of
individual galaxies. The high spatial resolution data that can now be
obtained with the Hubble Space Telescope (HST) allows the existing
evidence to be strengthened considerably. The present paper is part of
a new HST study of the quiescent galaxy M32, in which the presence of
a BH has long been suspected based on the steep central rotation
velocity gradient and nuclear peak in the velocity dispersion seen in
ground-based data (e.g., Tonry 1987; van der Marel \etal 1994a;
Bender, Kormendy \& Dehnen 1996). The main results of our project were
summarized, and discussed in the context of other recent work, in van
der Marel \etal (1997a). The acquisition and reduction of the stellar
kinematical HST data were described in van der Marel, de Zeeuw \& Rix
(1997b; hereafter Paper~I). Here we present new dynamical models that
we have used to interpret the combined HST and ground-based data.

Two types of self-consistent\footnote{We use the term
`self-consistent' for models in which the luminous mass density is in
equilibrium in the combined gravitational potential due to the
luminous mass density and some (known) dark matter density. This
definition is broader than the traditional one, which excludes dark
matter.} dynamical models have been constructed previously to
interpret the ground-based data for M32. Dressler \& Richstone (1988)
and Richstone, Bower \& Dressler (1990) used a method based on
Schwarzschild's (1979) technique, in which individual orbits are
calculated and superposed, to provide a self-consistent model that
fits a given set of data. These `maximum entropy' (Richstone \&
Tremaine 1988) models could fit the (then available) data only by
invoking the presence of a central dark mass of $(0.7$--$8) \times
10^6 M_{\odot}$. The models are general in the sense that they make no
assumptions about the dynamical structure of the galaxy. However, a
drawback was that only spherical geometry was considered. Even though
the models can be made to rotate, it remains unclear what systematic
errors are introduced when they are applied to a flattened (E3) galaxy
like M32. An alternative approach has been to construct axisymmetric
models with phase-space distribution functions (DFs) that depend only
on the two classical integrals of motion, $f=f(E,L_z)$, where $E$ is
the binding energy and $L_z$ is the angular momentum component along
the symmetry axis, both per unit mass. These models properly take
flattening and rotation into account. To fit the M32 data, they
require the presence of a central dark mass between $1.8 \times 10^6
M_{\odot}$ (van der Marel \etal 1994b; Qian \etal 1995; Dehnen 1995)
and $3 \times 10^6 M_{\odot}$ (Bender, Kormendy \& Dehnen 1996). The
disadvantage of these models is that they have a special dynamical
structure. The velocity dispersions in the meridional plane are
isotropic, $\sigma_r = \sigma_{\theta}$, which might not be the case
in M32. However, the models do fit the observed line-of-sight velocity
profile (VP) shapes without invoking freely adjustable parameters,
which provides some reason to believe that the M32 DF may not be too
different from the form $f(E,L_z)$.

The previous work on M32 has shown that models with a BH can fit the
ground-based data, but the modeling has not been general enough to
demonstrate that a BH is required.\footnote{Dressler \& Richstone
(1988) and Richstone, Bower \& Dressler (1990) argued that their data
could not be fit by any spherical model without a BH, but we show in
Figure~\ref{f:DRfit} below that their data can be fit by an
axisymmetric model without a BH.} In fact, the spatial resolution of
the ground-based data might have been insufficient for this to be the
case (see Appendix~A for a discussion of this issue). It has certainly
not been sufficient to rule out a cluster of dark objects (as opposed
to a central BH) on the basis of theoretical arguments; Goodman \& Lee
(1989) showed that this requires a resolution of $\lta 0.1''$. Our new
HST data of the nuclear region of M32 were obtained with the HST Faint
Object Spectrograph (FOS) through square apertures of $0.068''$ and
$0.191''$, respectively, yielding the highest spatial resolution
stellar kinematical data for M32 obtained to date. The results (see,
e.g., Figure~\ref{f:HSTBHpred} below) show a steeper rotation curve
and higher central velocity dispersion than the best ground-based M32
data. The primary goals of our project are to determine whether these
new HST data rigorously rule out models without any dark mass, and to
what extent they constrain the mass and size of the dark object in
M32.

To obtain constraints on the presence of a dark object that are least
dependent on a priori assumptions about the DF, we need to compare the
HST data not only to the predictions of axisymmetric $f(E,L_z)$
models, but also to the predictions of models with a fully general
dynamical structure. Orbit superposition techniques provide the most
straightforward approach to construct such models. However, orbit
superposition is more difficult to implement for the axisymmetric case
than for the spherical case: the orbits are not planar and typically
possess an additional integral of motion, so that the orbit library
must sample three rather than two integrals of motion. Furthermore, a
larger parameter space must be explored because of the unknown
inclination angle. This implies that larger amounts of CPU time and
computer memory are required. However, other than that, there are no
reasons why such models would be infeasible. Motivated by the
increased speed and memory capacity of computers, we therefore
developed a technique to construct fully general axisymmetric orbit
superposition models, that fit any given number of observed
photometric and kinematic constraints. Independent software
implementations were written by H.-W.R, N.C. and R.v.d.M. Our
technique may be viewed as the axisymmetric generalization of the
spherical modeling used by Richstone and collaborators, with the
important additional feature that we calculate VP shapes and include
an arbitrary number of Gauss-Hermite moments in the fit. We take into
account the error on each observational constraint to obtain an
objective $\chi^2$ measure for the quality-of-fit. Our basic algorithm
is described in Rix \etal (1997; hereafter R97) and summarized in de
Zeeuw (1997). R97 provide an application to the spherical geometry;
Cretton \etal (1997; hereafter C97) present the extension to the
axisymmetric case. Here we summarize the main steps of the
axisymmetric algorithm briefly, and focus on the application to
M32. The resulting models are the most general yet constructed for
M32.

The paper is organized as follows. In Section~\ref{s:massmodel} we
discuss our parametrizations for the stellar mass density and for the
potential of the dark object. In Section~\ref{s:DFpred} we describe
the construction of models with $f(E,L_z)$ DFs, and in
Section~\ref{s:datacomp} we compare the predictions of these models,
both with and without BHs, to the kinematical data. In
Section~\ref{s:technique} we outline the orbit-superposition technique
for constructing models with a fully general dynamical structure, and
in Section~\ref{s:results} we compare the predictions of these models
to the data. We construct models with an extended dark object in
Section~\ref{s:extfit}. We summarize and discuss our main conclusions
in Section~\ref{s:conc}. Readers interested primarily in the results
of our models may wish to skip Sections~\ref{s:massmodel},
\ref{s:DFpred} and~\ref{s:technique}.

\section{Mass density and potential}
\label{s:massmodel}

We adopt a parametrized form for the axisymmetric mass density of M32:
\begin{equation}
  \rho(R,z) = \rho_0 \> (m/b)^{\alpha} \> [1+(m/b)^2]^{\beta} 
                                       \> [1+(m/c)^2]^{\gamma} , \qquad
   m^2 \equiv R^2 + (z/q)^2 .
\label{massdens}
\end{equation}
The mass density $\rho$ is related to the luminosity density $j$
according to $\rho = \Upsilon j$, where $\Upsilon$ is the average
mass-to-light ratio of the stellar population (hereafter given in
solar V-band units). Both $\Upsilon$ and the intrinsic axial ratio $q$
are assumed to be constant (as a function of radius). The projected
axial ratio $q_p$ is determined by the inclination $i$ according to
$q_p^2 = \cos^2 i + q^2 \sin^2 i$. The parameters $\Upsilon$ and $i$
can be freely specified; all other parameters are determined by
fitting to the available M32 surface photometry.

The highest spatial resolution surface photometry available for M32 is
that presented by Lauer \etal (1992), based on pre-COSTAR HST/WFPC
images. Their measurements extend to $\sim 4''$ from the nucleus. At
larger radii ground-based data are available from Kent (1987) and
Peletier (1993). Figure~\ref{f:surfbr} shows the major axis surface
brightness measurements from these sources. The solid curve shows the
surface brightness profile predicted by our model, for
$\alpha=-1.435$, $\beta = -0.423$, $\gamma = -1.298$, $b=0.55''$,
$c=102.0''$, $q_p=0.73$, $\rho_0 = j_0 \Upsilon M_{\odot} /
L_{\odot,V}$, $j_0 = 0.463 \times 10^5 (q_p/q) L_{\odot,V} \pc^{-3}$,
and an assumed distance of $0.7 \Mpc$. The factor
$[1+(m/c)^2]^{\gamma}$ in equation~(\ref{massdens}) ensures that the
model has finite mass, and that it provides an adequate fit to the
observed surface brightness profile out to $\gta 100''$. Apart from
this factor, the model is identical to that used by van der Marel
\etal (1994b) and Qian \etal (1995) (dashed curve in
Figure~\ref{f:surfbr}).

Our model for the mass density is somewhat less general than that used
by Dehnen (1995), who deprojected the surface photometry in an
unparametrized manner. His approach avoids possible biases resulting
from the choice of an {\it ad hoc} parametrization (Merritt \&
Tremblay 1994; Gebhardt \etal 1996). It also allows the axial ratio of
M32 to vary with radius. The observed axial ratio is very close to
constant at $q_p = 0.73$ in the central $\sim 10''$, but increases
slowly to $0.86$ at $\sim 100''$. Even though our model does not
reproduce this modest variation, overall it provides an excellent fit,
and is fully adequate for a study of the nuclear dynamics. The
uncertainties in the interpretation of the kinematic data for the
center of M32 are due almost entirely to our ignorance of the
dynamical structure of M32. The uncertainties introduced by errors in
the brightness profile or by the non-uniqueness of the deprojection
are relatively minor (van den Bosch 1997). The effect of possible
triaxiality is more difficult to assess, but we will argue in
Section~\ref{ss:triaxiality} that triaxiality is unlikely to modify
any of the major conclusions of our paper.

The gravitational potential is assumed to be $\Psi = \Psi_{\rm lum} +
\Psi_{\rm dark}$, where $\Psi_{\rm lum}$ is the potential generated by
the luminous matter with mass density~(\ref{massdens}), and $\Psi_{\rm
dark}$ allows for the possibility of a massive dark object in the
nucleus. We assume the latter to be
\begin{equation}
   \Psi_{\rm dark} = G \Mdark (r^2 + \epsilon^2)^{-1/2} ,
\label{potent}
\end{equation}
which is the potential generated by a cluster with a Plummer model
mass density (e.g., Binney \& Tremaine 1987). For $\epsilon=0$ one
obtains the case of a dark nuclear point mass, i.e., a nuclear BH. We
do not include the potential of a possible dark halo around M32. There
are no (strong) observational constraints on the possible presence and
characteristics of such a dark halo, and even if present, it will not
affect the stellar kinematics near the nucleus of M32.

\section{Construction of two-integral models}
\label{s:DFpred}

The regular orbits in general axisymmetric potentials are
characterized by three integrals of motion, the binding energy $E =
\Psi - {1\over2}v^2$, the component of the angular momentum around the
symmetry axis $L_z = Rv_{\phi}$, and a non-classical, or effective,
third integral $I_3$ (Ollongren 1962; Richstone 1982; Binney \&
Tremaine 1987). In any given axisymmetric potential there is an
infinity of DFs $f(E,L_z,I_3)$ that generate a given axisymmetric
mass density $\rho(R,z)$. Such models are difficult to construct,
primarily because the third integral cannot generally be expressed
explicitly in terms of the phase-space coordinates. However, for any
mass density $\rho(R,z)$ there is exactly one DF that is even in
$L_z$, and does not depend on $I_3$. This unique even `two-integral'
DF, $f_{\rm e}(E,L_z)$, provides a useful low-order approximation to
any axisymmetric model, and has the convenient property that many
physical quantities, including the DF itself, can be calculated
semi-analytically. We study models of this type for M32 because they
have successfully reproduced ground-based M32 data, and because they
provide a useful guide for the interpretation of more general
three-integral models, which are discussed in
Sections~\ref{s:technique} and~\ref{s:results}.

To calculate the $f_{\rm e}(E,L_z)$ DFs for our models we have used a
combination of the techniques described in Qian \etal (1995) and
Dehnen (1995), even though either technique by itself could have been
used to get the same result (in fact, yet another technique to address
this problem is described in C97). Initially, four radial regimes are
considered: $m \ll b$; $m \approx b$; $m \approx c$; and $m \gg c$,
with $m$ and $b \ll c$ as defined in equation~(\ref{massdens}). In
these regimes the mass density is approximately: $\rho \propto
m^\alpha$; $\rho \propto (m/b)^{\alpha}
\> [1+(m/b)^2]^{\beta}$; $\rho \propto (m/b)^{\alpha+2\beta}
[1+(m/c)^2]^{\gamma}$; and $\rho \propto m^{\alpha+2\beta+2\gamma}$,
respectively. For each of these mass densities $f_{\rm e}(E,L_z)$ can
be calculated with the technique and software of Qian \etal
(1995). The DFs for the four regimes are then smoothly patched
together in energy, to yield an approximation to the full DF.  This
approximation is then used as the starting point for Lucy-Richardson
iteration as described in Dehnen (1995). This yields the DF $f_{\rm
e}(E,L_z)$, reproducing the model mass density to $\lta 0.3$ per cent
RMS.

The total DF is the sum of the part $f_{\rm e}$ that is even in $L_z$,
and the part $f_{\rm o}$ that is odd in $L_z$. In principle, $f_{\rm
o}$ is determined completely by the mean streaming velocities $\langle
v_{\phi} \rangle (R,z)$, but these are not determined well enough by
the data to make an inversion practicable. Instead, therefore, $f_{\rm
o}$ can be freely specified so as to best fit the data, with the only
constraint that the total DF should be positive definite. None of the
main conclusions of our paper depend sensitively on the particular
parametrization used for $f_{\rm o}$, so we restrict ourselves here to
a simple choice (Section~\ref{ss:BHfit}). Once the complete DF is
known, the projected line-of-sight VPs can be calculated for any
particular observational setup as in Qian \etal (1995). From the VPs,
predictions can be calculated for the observable kinematical
quantities.

\section{Predictions of two-integral models}
\label{s:datacomp}

\subsection{Data-model comparison}
\label{ss:BHfit}

Figure~\ref{f:HSTBHpred} shows the HST/FOS data presented in Paper~I,
obtained with the apertures `0.1-PAIR' ($0.068''$ square) and
`0.25-PAIR' ($0.191''$ square). The figure also shows the highest
available spatial resolution ground-based data, obtained by van der
Marel \etal (1994a) with the William Herschel Telescope (WHT), and by
Bender, Kormendy \& Dehnen (1996) with the Canada-France Hawaii
Telescope (CFHT). The spatial resolution for these ground-based
observations is roughly $0.9''$ and $0.5''$ FWHM, respectively.  The
predictions of $f(E,L_z)$ models for the WHT and CFHT data have
already been discussed in detail by previous authors, and we therefore
focus here on a comparison of $f(E,L_z)$ models to the new HST data.

The curves in Figure~\ref{f:HSTBHpred} are the predictions of edge-on
$f(E,L_z)$ models with dark nuclear {\it point masses} (i.e.,
BHs). These models have an intrinsic axial ratio $q=0.73$. The dark
mass only influences the kinematical predictions in the central few
arcsec, and the mass-to-light ratio $\Upsilon$ was therefore chosen to
fit the normalization of the kinematical data at larger radii. A good
fit to the WHT data between $\sim 5''$ and $\sim 12''$ is obtained
with $\Upsilon = 2.51$. Predictions were calculated for each
individual observation, taking into account the aperture position,
aperture size and PSF for the HST data as given in Paper~I. Connecting
curves in the figure were drawn to guide the eye. It proved sufficient
to study only models with a very simple odd part $f_{\rm o}$, namely
those that produce a total DF in which at every $(E,\vert L_z\vert)$ a
fraction $F$ of the stars has $L_z > 0$, and a fraction $(1-F)$ has
$L_z < 0$.  For each $\Mdark$, the fraction $F \leq 1$ of stars with
$L_z > 0$ in the model was chosen to optimize the $\chi^2$ of the fit
to the rotation curve. The displayed models {\it with} a BH fit the
rotation curve well, more or less independent of $\Mdark$; models with
higher $\Mdark$ require smaller~$F$. The models differ primarily in
their predictions for the velocity dispersions. The observed trend of
increasing velocity dispersion towards the nucleus is successfully
reproduced by models with a nuclear point mass of $\Mdark \approx (3
\pm 1) \times 10^6 \Msun$. The model without a BH predicts a roughly constant
velocity dispersion with radius, and is strongly ruled out. In fact,
this model also fails to fit the observed rotation velocity gradient
in the central arcsec. The displayed model for the no-BH case is
maximally rotating ($F=1$), and it is thus not possible to improve
this by choosing a more general form for the odd part of the DF.

For a quantitative analysis of the best-fitting $\Mdark$, we define a
$\chi^2$ statistic that measures the quality of the model fit to the
observed HST velocity dispersions:
\begin{equation}
  \chi_{\sigma}^2 \equiv \sum_{i=1}^{N}
        \left ( {{\sigma_{\rm model}-\sigma_{\rm obs}}
                   \over{\Delta \sigma_{\rm obs}}}
        \right )^2   , \qquad
  \chi_{\sigma,{\rm RMS}} \equiv [\chi_{\sigma}^2/N]^{1/2} .
\label{chisigdef}
\end{equation}
Figure~\ref{f:chisqBH} shows the relative RMS residual
$\chi_{\sigma,{\rm RMS}}$ as function of $\Mdark$. The best fit to the
HST dispersions is obtained for $\Mdark = (2.7 \pm 0.3) \times 10^6
\Msun$. The quoted error is a formal $1\sigma$ error based on the
assumption of Gaussian statistics (probably an underestimate, as there
is some hint for systematic errors in the data in addition to random
errors). An alternative way of estimating $\Mdark$ is to model the
average of the four data points within $0.1''$ from the center:
$\sigma_{r \leq 0.1''} = (126 \pm 10) \kms$, where the $1\sigma$ error
is based on the scatter between the data points. Figure~\ref{f:sigcen}
shows the observed $\sigma_{r \leq 0.1''}$ as a hatched region; 
a solid curve shows the predicted value as a function of $\Mdark$. The
predictions fall in the observed range for $\Mdark = (3.4 \pm 0.9)
\times 10^6 \Msun$.

The inclination of M32 cannot be derived from the observed photometry
and is therefore a free parameter in the modeling. However, the
predictions of $f(E,L_z)$ models are rather insensitive to the assumed
inclination (van der Marel \etal 1994b). This was verified by also
calculating the predictions of inclined models with $i=55^{\circ}$,
which have an intrinsic axial ratio $q=0.55$. A mass-to-light ratio
$\Upsilon=2.55$ was adopted, so that at large radii one obtains the
same RMS projected velocity on the intermediate axis (between the
major and minor axes) as for an edge-on model. On the major axis the
$i=55^{\circ}$ model then predicts a slightly higher RMS velocity than
the edge-on model, resulting in a $\sim 10$ per cent smaller
best-fitting $\Mdark$. Apart from this, the conclusions from the
inclined models were found to be identical to those for the edge-on
models (cf.~Figure~\ref{f:chisqsurf} below).\looseness=-2

\subsection{Is the M32 distribution function of the form $f(E,L_z)$?}
\label{ss:DFform}

The $f(E,L_z)$ models that fit the HST data can only be correct if
they also fit the ground-based WHT and CFHT data. One may define
similar $\chi^2$ quantities as for the HST data, to determine the
best-fitting $\Mdark$ for either of these data
sets. Figures~\ref{f:chisqBH} and~\ref{f:sigcen} show the relative RMS
residual $\chi_{\sigma,{\rm RMS}}$ of the fit to all the dispersion
measurements, and the average dispersion $\sigma_{r
\leq 0.1''}$ of the dispersion measurements centered with $0.1''$ from
the nucleus. For the WHT data, $\chi_{\sigma,{\rm RMS}}$ is minimized
for $\Mdark = (1.46 \pm 0.03) \times 10^6 \Msun$, whereas for the CFHT
data it is minimized for $\Mdark = (2.8 \pm 0.2) \times 10^6
\Msun$. The central velocity dispersion measured with the WHT is best 
fit with $\Mdark = (1.94 \pm 0.05) \times 10^6 \Msun$, whereas for the
CFHT data it is best fit with $\Mdark = (2.1 \pm 0.2) \times 10^6
\Msun$. These results are roughly consistent with those of previous
authors. Van der Marel \etal (1994b), Qian \etal (1995) and Dehnen
(1995) found $\Mdark \approx (1.8 \pm 0.3) \times 10^6 \Msun$ for the
best fitting edge-on $f(E,L_z)$ model to the ground-based WHT data,
while Bender, Kormendy \& Dehnen obtained a best fit to their higher
spatial resolution CFHT data with $\Mdark \approx (3.0 \pm 0.5) \times
10^6 \Msun$. The latter value is somewhat higher than the one we find
here, because it was chosen only to provide a good fit to the CFHT
rotation curve; it does not fit the CFHT velocity dispersions very
well.
 
These results indicate that, under the assumption of an $f(E,L_z)$ DF,
the different observations cannot all be fit simultaneously with the
same $\Mdark$, even after accounting for the different observational
setups. The lowest spatial resolution WHT data require a significantly
lower $\Mdark$ than the highest spatial resolution HST data. This
implies that M32 has a DF that is not of the form $f=f(E,L_z)$.

\section{Construction of three-integral models}
\label{s:technique}

To construct more general three-integral models for M32, we extended
Schwarzschild's orbit superposition algorithm. Its basic structure is
to calculate an orbit library that samples integral space in some
complete and uniform way, to store the time-averaged intrinsic and
projected properties of the orbits, and to search for the weighted
superposition of orbits that best fits the observed kinematics, while
reproducing the mass density $\rho(R,z)$ for self-consistency. Here we
summarize the main steps, with emphasis on those aspects that are
unique to the M32 application. Complete descriptions of the technique
are given in R97 and C97.

We sample integral space with an $(\Rc,\eta,w)$ grid. The quantity
$\Rc(E)$ is the radius of the circular orbit in the equatorial plane
with energy $E$. Its angular momentum, $L_{\rm max}(E)$, is the
maximum angular momentum at the given energy. We define $\eta(E,L_z)
\equiv L_z / L_{\rm max}(E)$. For fixed $(\Rc,\eta)$, the position of
a star in the meridional ($R,z)$ plane is restricted to the region
bounded by the `zero-velocity-curve' (ZVC), defined by the equation $E
= \Psi_{\rm eff}$, where $\Psi_{\rm eff} = \Psi - {1\over2} L_z^2/R^2$
is the `effective gravitational potential' (Binney \& Tremaine
1987). We parametrize the third integral at each $(\Rc,\eta)$ using an
angle $w$, which fixes the position at which an orbit touches the ZVC
(cf.~Figure~\ref{f:zvcplot}).

The quantity $\Rc$ was sampled using 20 logarithmically spaced values
between $R_{\rm c,min} = 6.12 \times 10^{-4}$ arcsec, and $R_{\rm
c,max} = 7.55 \times 10^{3}$ arcsec. This range of radii contains all
but a fraction $10^{-4}$ of the stellar mass of M32. The quantity
$\eta$ was sampled using an `open' grid (in the same sense that
numerical quadrature formulae can be open or closed, e.g., Press
\etal 1992) of $N_{\eta}=14$ values, spaced linearly between 
$-1$ and $1$, i.e., $\eta_i = -1 + (2i-1)/N_{\eta}$, for
$i=1,\ldots,N_{\eta}$. The quantity $w$ was sampled using an `open'
grid of 7 values, spaced linearly between 0 and $w_{\rm th}$. Here
$w_{\rm th}$ is the angle $w$ for the `thin tube' orbit at the given
$(E,L_z)$ (see Figure~\ref{f:zvcplot}). The special values $\eta = 0$
and $\eta = \pm 1$ (meridional plane and circular orbits) and the
special values $w= 0$ and $w = w_{\rm th}$ (equatorial plane and thin
tube orbits) are presumed to be represented by their closest neighbors
on the grid, but are not included explicitly.

An orbit was integrated for each $(\Rc,\eta,w)$ combination, starting
with $v_R = v_z = 0$ from the ZVC. The integration time was 200 times
the period of the circular orbit at the same energy. This is
sufficient to properly sample the phase-space trajectory for the large
majority of orbits, although it exceeds the Hubble time only at radii
$\gta 100''$. The integrations yield the `orbital phase-space density'
for each orbit, as described in C97. These were binned onto: (i) an
$(r,\theta)$ grid in the meridional plane; (ii) an $(r',\theta')$ grid
on the projected $(x,y)$ plane of the sky; and (iii) several Cartesian
$(x,y,v)$ cubes, with $v$ the line-of-sight velocity. The first two
grids were chosen logarithmic in $r,r'$, with identical bins to those
used for $\Rc$, and linear in $\theta,\theta' \in
[0,{\pi\over2}]$. The $(x,y,v)$ cubes were centered on $(0,0,0)$, with
$211 \times 211$ square spatial cells, and 91 velocity bins of $15
\kms$. Spatial cell sizes were adopted of $0.025''$, $0.08''$ and $0.5''$,
respectively. The $(x,y,v)$ cubes were used to calculate, for each
orbit, the predicted line-of-sight velocity histograms for all
positions and setups for which kinematical data are available, taking
into account the observational point-spread-functions (PSFs) and
aperture positions, orientations and sizes, as described in C97.

Construction of a model consists of finding a weighted superposition
of the orbits in the library that reproduces two sets of constraints:
\begin{itemize}
\item Self-consistency constraints. The model should reproduce the
masses predicted by the luminous density $\rho(R,z)$
(Section~\ref{s:massmodel}), for each cell of the meridional
$(r,\theta)$ grid, for each cell of the projected $(r',\theta')$ grid,
and for each aperture on the sky for which there is data\footnote{In
theory, it is sufficient to fit only the meridional plane
masses. Projected masses are then fit automatically. In practice this
is not exactly the case, because of discretization. Projected masses
were therefore included as separate constraints.}.
\item Kinematical constraints. The model should reproduce the observed
kinematics of the galaxy, including VP shapes. We chose to express
this as a set of linear constraints on the Gauss-Hermite moments $h_i$
of the VPs (R97).
\end{itemize}
Smoothness of the solutions in integral space may be enforced by
adding extra regularization constraints (see
Section~\ref{ss:smoothmod} below). The quality of the fit to the
combined constraints can be measured through a $\chi^2$ quantity
(R97). The assessment of the fit to the kinematical constraints
includes the observational errors. In principle, one would like to fit
the self-consistency constraints with machine precision. In practice
this is unfeasible, because the projected mass constraints are not
independent from each other (aperture positions for different data
sets partly overlap) and from the meridional plane mass
constraints. It was found that models with no kinematical constraints
could at best fit the masses with a fractional error of $\sim 5 \times
10^{-3}$. Motivated by this, fractional `errors' of this size were
assigned to all the masses in the self-consistency
constraints\footnote{In principle one would like to include the
observational surface brightness errors in the
analysis. Unfortunately, this requires the exploration of a large set
of three-dimensional mass densities (that all fit the surface
photometry to within the errors), which is prohibitively
time-consuming. However, the observational errors in the surface
brightness are small enough that they are not believed to influence
the conclusions of our paper.}. As described in R97 and C97, we use
the NNLS routine of Lawson \& Hanson (1974) to determine the
combination of non-negative orbital occupancies (which need not be
unique) that minimizes the combined $\chi^2$.

The model predictions have a finite numerical accuracy, due to, e.g.,
gridding and discretization. Tests show that the numerical errors in
the predicted kinematics of our method are $\lta 2 \kms$ for the
rotation velocities and velocity dispersions, and $\lta 0.01$ in the
Gauss-Hermite moments (C97). Numerical errors of this magnitude have
only an insignificant effect on the data-model comparison (see
Appendix~A).

\section{Predictions of three-integral models}
\label{s:results}

\subsection{Implementation}

We have studied three-integral models with dark central {\it
point}-masses, i.e., $\epsilon=0$ in equation~(\ref{potent}). There
are then three free model parameters: the inclination $i$, the
mass-to-light ratio $\Upsilon$, and the BH mass $\Mdark$. The
parameter space must be explored through separate sets of orbit
libraries\footnote{Only one orbit library needs to be calculated for
models with the same $\Mdark/\Upsilon$. The potentials of such models
are identical except for a normalization factor, and the orbits are
therefore identical except for a velocity scaling. Each
$(\Upsilon,\Mdark)$ combination does require a separate NNLS fit to
the constraints.}. As for the two-integral models, we have studied
only two, widely spaced, inclinations: $i=90^{\circ}$ and
$i=55^{\circ}$. For each inclination we have sampled the physically
interesting range of $(\Upsilon,\Mdark)$ combinations. For each
$(\Upsilon,\Mdark)$ combination we determined the orbital weights (and
hence the dynamical structure) that best fit the data, and the
corresponding goodness-of-fit quantity $\chi^2 (\Upsilon,\Mdark)$. All
available HST/FOS, CFHT and WHT data were included as kinematical
constraints on the models. Older kinematical data for M32 were not
included because of their lower spatial resolution and/or poorer sky
coverage. In total, each NNLS fit had 1960 orbits to fit 782
constraints: 366 self-consistency constraints, and 416 kinematical
constraints (for 86 positions on the projected plane of the
galaxy). We focus primarily
(Figures~{\ref{f:contour}}--{\ref{f:datafitsAtoH}}) on models without
additional constraints that enforce smoothness in integral-space. This
is a sufficient and conservative approach for addressing the primary
question of our paper: {\it which models are ruled out by the M32
data?} If the data cannot even be fit with an arbitrarily unsmooth DF,
they certainly cannot be fit with a smooth DF.

\subsection{Data-model comparison}
\label{ss:threedatacomp}

Figure~\ref{f:contour} shows the main result: contour plots of
$\chi^2(\Upsilon,\Mdark)$ for both inclinations that were studied. The
displayed $\chi^2$ measures the quality of the fit to the kinematical
constraints only; the actual NNLS fits were done to both the
kinematical and the self-consistency constraints, but contour plots of
the total $\chi^2$ look similar. The overall minimum $\chi^2$ values
are obtained for: $\Upsilon=2.1$ and $\Mdark = 3.4
\times 10^6 \Msun$ for $i=90^{\circ}$; and $\Upsilon = 2.0$ and
$\Mdark = 3.2 \times 10^6 \Msun$ for $i=55^{\circ}$. 

Figure~\ref{f:bestfits} compares the kinematical predictions of the
best-fitting edge-on and $i=55^{\circ}$ models to the data. Two
problems with the data must be taken into account when assessing the
quality of the fit. First, the HST velocity dispersions show a scatter
between some neighboring points that is much larger than the formal
errors, most likely due to some unknown systematic effect. The models
cannot be expected to reproduce this. Second, the CFHT rotation
velocities at radii $\gta 0.5''$ exceed the WHT measurements by an
amount which cannot be attributed to differences in spatial
resolution, but must be due to some unknown systematic error in either
of the two data sets. The WHT data have the smallest error bars, and
therefore receive most weight in the NNLS fit. As a result, the models
tend to underpredict the CFHT rotation curve.

These systematic problems with the data preclude the use of $\chi^2$
as a meaningful statistic to assess which models provide an acceptable
fit: if the observations themselves are not mutually consistent, then
clearly no model can be statistically consistent with all of
them. Although the use of any statistical test is suspect in the
presence of systematic errors, one may still assign confidence regions
on the model parameters by using the {\it relative likelihood}
statistic $\Delta \chi^2 \equiv \chi^2 - \chi^2_{\rm min}$. This
statistic merely measures which parameter combinations provide an
equally good (or bad) fit as the one(s) that yield the minimum
$\chi^2$. If we assume that the observational errors are normally
distributed (which, as mentioned, is likely to be an
oversimplification), then $\Delta \chi^2$ follows a $\chi^2$
probability distribution with the number of degrees of freedom equal
to the number of model parameters (Press \etal 1992)\footnote{A more
robust way to incorporate the effects of {\it random} errors in the
assignment of confidence bands would be to use `bootstrapping', in
which one directly calculates the statistical distribution of models
parameters by finding the best-fit parameter combinations for
different `realizations' of the data set.  Unfortunately, this is
computationally infeasible in the present context: even the analysis
of the single (available) data set for M32 already takes weeks of CPU
time on a high-end workstation.}.

The best-fitting edge-on model in Figure~\ref{f:bestfits} has $\chi^2
= 690.0$ while the best-fitting $i=55^{\circ}$ model has $\chi^2 =
602.5$, both for $N=416$ degrees of freedom. The fact that $\chi^2 >
N$, even for these optimum fits, is due primarily to the systematic
errors in the data. To the eye, the models appear to fit the data as
well as could be hoped for. The $\chi^2$ values do indicate that the
$i=55^{\circ}$ model provides a significantly better fit than the
edge-on model, implying that M32 is not seen edge-on. However, the
results presented here do not allow us to derive the actual
inclination of M32. That would require a detailed study of the entire
range of possible inclinations, which would be more
computer-intensive. The important conclusion in the present context is
that the topology of the $\chi^2$ contours in Figure~\ref{f:contour}
is virtually identical for both inclinations: the allowed range for
$\Mdark$ is therefore uninfluenced by our ignorance of the true
inclination of M32.

The $\Delta \chi^2$ statistic was used to assign confidence values to
the contours in Figure~\ref{f:contour}. At the $68.3$ per cent
confidence level ($1\sigma$ for a Gaussian probability distribution),
the allowed $\Mdark$ fall in the range $(3.2$--$3.5) \times 10^6
\Msun$ for $i=90^{\circ}$, and in the range $(3.1$--$3.4) \times 10^6
\Msun$ for $i=55^{\circ}$. At the $99.73$ per cent confidence level
($3\sigma$ for a Gaussian probability distribution), they fall in the
ranges $(2.5$--$3.7) \times 10^6 \Msun$ and $(2.3$--$3.9) \times 10^6
\Msun$, respectively. In reality, small numerical errors in the
models might have distorted the $\chi^2$ contours. We address this
issue in Appendix~A. Any numerical errors are small enough that they
have no influence on our conclusion that models without a dark mass
are firmly ruled out. However, the possibility of small numerical
errors does increase the confidence bands on $\Mdark$. Based on the
analysis in Appendix~A, we conclude that $\Mdark = (3.4 \pm 0.7)
\times 10^6 \Msun$ at $68.3$ per cent confidence, and $\Mdark = (3.4 \pm 1.6)
\times 10^6 \Msun$ at $99.73$ per cent confidence. These estimates
take into account both the observational errors in the data and
possible numerical errors in the models, and are valid for both
inclinations that were studied.

Figure~\ref{f:datafitsAtoH} compares the model predictions to the
observed rotation velocities and velocity dispersions for all the
models labeled in Figure~\ref{f:contour}. Models~C\&G are the overall
best fits for the two inclinations. Models~B\&F and~D\&H are
(approximately) the best-fitting models for $\Mdark=1.9$ and $5.4
\times 10^6 \Msun$, respectively. The latter models are marginally
ruled out at the $>99$\% confidence level (cf.~the above discussion),
although to the eye they do appear to reproduce the main features of
the data.  They differ from the overall best-fitting models primarily
in their predictions for the HST velocity dispersions. The differences
in the predictions for the ground-based data are smaller (and
invisible to the eye in Figure~\ref{f:datafitsAtoH}), but nonetheless
more statistically significant because of the smaller error bars for
these data. Models~A\&E, the best fits without a central dark mass, are
indisputably ruled out. The main problem for these models is to fit
the central peak in the velocity dispersion. They come rather close to
fitting the WHT observations, and predict a central dispersion of
$\sim 84 \kms$.  However, the models without a dark mass fail to
reproduce the higher central dispersion of $\sim 91 \pm 2 \kms$
measured with the CFHT (although still only marginally), and don't
even come close to reproducing the HST dispersions, which exceed $100
\kms$ in the central $0.1''$.

\subsection{Smooth solutions}
\label{ss:smoothmod}

The top row in Figure~\ref{f:intspace} illustrates the orbital
occupancies in integral space for the best-fitting edge-on model
(model~C). Only a small fraction of the orbits in the library is used
to fit the constraints, while the remainder receives zero weight.
This yields an equilibrium solution of the collisionless Boltzmann
equation, but is not physically plausible.

Smoothness of the solutions in integral space can be enforced by
adding linear regularization constraints to the problem (Zhao 1996;
C97). We have explored this only in an {\it ad hoc} way, merely to be
able to assess the effect of smoothness constraints on the resulting
fit to the data. A model is defined as a set of masses $m(\Rc,\eta,w)$
in integral space. For each point that is not on the boundary of the
$(\Rc,\eta,w)$ grid, we measure the smoothness of the model (Press
\etal 1992; eq.~[18.5.10]) through the second order divided differences
(in each of the three variables, assuming for simplicity that the
distances between adjacent grid points are equal in all directions) of
the function $m(\Rc,\eta,w)/m_0(\Rc)$. The function $m_0(\Rc)$ is a
rough approximation to the energy dependence of the model, obtained,
e.g., by studying the spherical isotropic limit of the given mass
density. The regularization constraints are then that the divided
differences should equal $0 \pm \Delta$, where the `error' $\Delta$
determines the amount of smoothing. Models with $\Delta \rightarrow
\infty$ have no smoothing, while models with $\Delta \rightarrow 0$
force $m(\Rc,\eta,w)/m_0(\Rc)$ to be a linear function on the
$(\Rc,\eta,w)$ grid.

The second and third rows in Figure~\ref{f:intspace} show the integral
space for model~C with the addition of either a modest ($\Delta=5$) or
a large ($\Delta = 0.2$) amount of regularization in the NNLS fit,
respectively. As the bottom panels show, the price paid for the
increased smoothness is a somewhat poorer fit to the data. However,
the fits are still quite good. This demonstrates that the good fits to
the data shown in Figure~\ref{f:bestfits} are not primarily the result
of the use of implausible distributions in integral space. These
distributions result from the numerical properties of the problem, but
there also exist smooth solutions which provide similar fits.

\subsection{Dynamical Structure}
\label{ss:dynstruc}

Figure~\ref{f:moments} illustrates the dynamical structure of the
edge-on models~A--D and the $i=55^{\circ}$ models~E--H. By contrast to
the models with a BH, the models~A\&E without a BH invoke a large
amount of radial motion in the central arcsec to produce a peak in the
observed velocity dispersions (cf.~Binney \& Mamon 1982). The maximum
allowed radial anisotropy in the models is determined by the observed
rotation velocities in M32, because dynamical models predict lower
values of $V/\sigma$ when they are more radially anisotropic
(Richstone, Bower \& Dressler 1990; de Bruijne, van der Marel \& de
Zeeuw 1996). Figure~\ref{f:datafitsAtoH} shows that the allowed radial
anisotropy is by far insufficient to fit the observed peak in the
velocity dispersion profile without invoking a BH.

The models~B--D and~F--H, which represent the best fits for different
potentials, all have a similar dynamical structure in that they are
dominated by azimuthal motion. This is most pronounced for the
intrinsically flatter $i=55^{\circ}$ models. It is not known what
physical process could have produced this particular velocity
ellipsoid shape in M32. The rightmost panels in Figure~\ref{f:moments}
show the velocity moments for $f(E,L_z)$ models with the same
gravitational potentials as models~C\&G. The $f(E,L_z)$ models are
similar to the best-fitting three-integral models, in that they have
an excess of azimuthal motion. This is why they have been so
successful in fitting ground-based data, including available VP shape
parameters, and it shows that they provide a useful low-order
approximation to the dynamical structure of M32. However, the
$f(E,L_z)$ models have $\vrsqav=\vthsqav$ by definition, in contrast
to the inequality between $\vrsqav$ and $\vthsqav$ seen in the
best-fitting three-integral models. This explains the finding of
Section~\ref{ss:DFform} that $f(E,L_z)$ models cannot successfully
reproduce {\it all} observed features of the kinematical data.

The second velocity moments $\vrsqav$ and $\vthsqav$ can be combined
with the mixed moment $\langle v_r v_\theta \rangle$ to determine the
tilt of the velocity ellipsoid in the meridional plane. For all the
models A--D and E--H we found $\langle v_r v_\theta \rangle$ to be
small, and the velocity ellipsoids are more closely aligned with
spherical coordinate axes than with cylindrical coordinate axes (but
they are not perfectly aligned with either). This is not uncommon in
three-integral models for axisymmetric systems (e.g., Dejonghe \& de
Zeeuw 1988; Dehnen \& Gerhard 1993; de Zeeuw, Evans \& Schwarzschild
1996).

Figure~\ref{f:vsig} shows the quantity $\langle v_\phi \rangle \, / \,
\langle v^2 \rangle^{1/2}$ in the equatorial plane, for the models~C and~G.
The inclined, and intrinsically flatter, model~G rotates faster than
the edge-on model~C. The predictions for oblate isotropic rotator
models ($\sigma_r = \sigma_\theta = \sigma_\phi$) are shown for
comparison. The edge-on model~C rotates faster than an oblate
isotropic rotator model for radii $r \gta 0.5''$; the inclined model~G
rotates slower than the oblate isotropic rotator model.

We have not attempted to derive confidence bands on the dynamical
structure of M32. This is a much more difficult problem than the
derivation of confidence bands on the model parameters
$(\Mdark,\Upsilon)$, and is beyond the scope of the present paper.

\section{Models with an extended dark nuclear object}
\label{s:extfit}

The results in the previous section demonstrate that M32 must have a
massive dark object in its nucleus. To obtain a limit on the size of
this dark object, we have studied models in which it has a finite size
$\sim \epsilon$ (cf.~eq.~[\ref{potent}]).  Searching the parameter
space of three-integral models with different $\epsilon$ is extremely
computer-intensive. We have therefore restricted ourselves to
$f(E,L_z)$ models with extended dark objects. This is not likely to
bias our conclusions, because the best-fitting three-integral models
found in Section~\ref{s:results} are similar to two-integral models.

Figure~\ref{f:HSTextpr} shows the predictions of edge-on $f(E,L_z)$
models with extended nuclear dark objects, for two representative
values of $\Mdark$. As in the models of Section~\ref{ss:BHfit}, the
fraction $F$ of stars with $L_z > 0$ in each model was chosen to best
fit the rotation curve. By adjusting $F$, adequate fits to the
rotation velocities can be obtained for most relevant models. Hence,
only the velocity dispersions are shown in the figure. The predicted
dispersions are typically constant or decreasing towards the center
within the scale radius $\epsilon$. The HST data show a much higher
velocity dispersion in the center than further out. Hence, the
extension of any possible dark nuclear cluster cannot be large;
Figure~\ref{f:HSTextpr} suggests $\epsilon \lta 0.1''$.
Figure~\ref{f:chisqsurf} shows a contour plot of the quantity
$\chi_{\sigma}^2$ (eq.~[\ref{chisigdef}]), measuring the quality of
the model fit to the HST dispersion measurements, both for the edge-on
case and for $i=55^{\circ}$. The best fit is obtained for $\epsilon =
0$ (the point mass case discussed previously). The formal $99.73$ per
cent confidence level (assuming Gaussian formal errors) rules out all
models with $\epsilon \gta 0.06''$, independent of the
inclination. The models with the largest $\epsilon$ must have a total
mass of at least $\Mdark \approx 4 \times 10^6 \Msun$.

At the distance of M32, $1'' = 3.39 \pc$. Hence, the upper limit on
the scale radius corresponds to $\epsilon = 0.20 \pc$. Combined with a
total mass in the cluster of $\Mdark \approx 4 \times 10^6 \Msun$,
this implies a central mass density of at least $\rho_0 = 1.1 \times
10^8 \Msun \pc^{-3}$. The half-mass radius of a Plummer model is
$r_{\rm h} = 1.30 \epsilon$. Hence, there must be $\sim 2 \times 10^6
\Msun$ inside $r \lta 0.078''$. The total V-band luminosity inside
this radius\footnote{This quantity does not depend sensitively on the
assumed density cusp slope at very small radii. Gebhardt \etal (1996)
infer a somewhat steeper slope for M32 than used here, but their model
has only 10 per cent more luminous mass inside $r \lta 0.078''$ than
ours.} is $1 \times 10^5 \Lsun$, implying a luminous mass of $2.5
\times 10^5 \Msun$. Hence, the ratio of the total mass to luminosity
inside this radius must be $\gta 22.5$.

The observed kinematics constrain only the amount of mass in the
system, not whether this mass is luminous or dark. One can therefore
fit the data equally well with models in which the average
mass-to-light ratio $\Upsilon$ of the stellar population increases
towards the nucleus, and in which there is no dark mass. We have not
explicitly constructed such models, but it is clear from the preceding
discussion that in such models $\Upsilon$ must rise from $\sim 2$ in
the main body of the galaxy to $\gta 20$ at $r \lta 0.1''$. Such a
drastic variation in mass-to-light ratio would imply a strong change
in the stellar population, accompanied by broad-band color
gradients. The size of these gradients depends on the actual stellar
population mix, which is unknown. However, one may use the properties
of main sequence stars as a guideline. For these, a change in
$\Upsilon$ from $2$ to $20$ implies color changes $\Delta(U-B) \approx
0.9$, $\Delta(B-V) \approx 0.6$, and $\Delta(V-I) \approx 0.8$ (using
the tables of stellar properties in Allen 1973). Such variations
between $0.1''$ and $1''$ should have been obvious in photometric
observations. However, neither subarcsec resolution ground-based
imaging (Lugger \etal 1992) nor pre-refurbishment HST imaging (Crane
\etal 1993) have revealed any significant color gradients in the
central arcsec of M32. Post-refurbishment HST observations (Lauer et
al., private communication) also do not show strong color
gradients. Thus, the nuclear mass concentration in M32 cannot be due
merely to a change in the mix of ordinary stars in the nuclear
region.\looseness=-2

The absence of observed color gradients does not exclude the
possibility of nuclear concentrations of brown dwarfs, white dwarfs,
neutron stars, or stellar-mass BHs in M32. However, at high densities,
such clusters of dark objects are not stable over a Hubble time. This
was discussed by Goodman \& Lee (1989), and their arguments were
updated and extended in van der Marel \etal (1997a). The latter paper
shows that the new HST limit on the density of dark material in M32
rules out all but the most implausible clusters, leaving a single
massive BH as the most likely interpretation of the data.

The kinematical predictions of our models depend on the assumed
Plummer form of the extended dark object. If it is a cluster of
collapsed objects, this distributed dark mass may itself be cusped
(e.g., Gerhard 1994). However, the limit on $\epsilon$ results from
the fact that the dispersion of stars with mass
density~(\ref{massdens}) in a Plummer potential does not have a
(strong) central peak. This property is common to many alternative
types of models, such as those of King (e.g., Binney \& Tremaine
1987), Hernquist (1990) and Jaffe (1983). These all produce constant
or decreasing dispersions inside their scale radius, as does the
Plummer model. Hence, the upper bound on $\epsilon$ derived here is
likely to be generic to most plausible density profiles for the
extended dark object. In addition, King, Jaffe and Hernquist models
are more centrally condensed than Plummer models, and would therefore
require dark clusters of even higher central densities to fit the
data.

\section{Conclusions and discussion}
\label{s:conc}

\subsection{Summary of results}

The main bottlenecks in proving the presence of nuclear BHs in
quiescent galaxies from stellar kinematical data have long been: (i)
the restricted spatial resolution of ground-based data; and (ii) lack
of sufficiently general dynamical models to rule out constant
mass-to-light ratio models beyond doubt. The HST now provides spectra
of superior spatial resolution. To fully exploit the potential of
these new data it is imperative to improve the modeling techniques
that have been used in the past decade. The situation is considerably
more complicated than for gas disks in (active) galaxies, where the
assumption of simple circular orbits is often adequate. Interpretation
of stellar kinematical data for flattened elliptical galaxies ideally
requires axisymmetric (or even better, triaxial) dynamical models with
completely general three-integral distribution functions. Such models
have not previously been constructed for any stellar kinematical BH
candidate galaxy. We therefore developed a technique for the
construction of such axisymmetric models, and used it to interpret our
HST data for M32.

To guide the construction and interpretation of the three-integral
models we first compared the new HST data to the predictions of
$f(E,L_z)$ models, which have been used extensively to interpret
ground-based M32 data. Such models have the advantage that the DF can
be calculated semi-analytically, but have the disadvantage of having a
special dynamical structure, with $\sigma_r=\sigma_{\theta}$
everywhere. There is no a priori reason why any galaxy should have
this property. However, the fact that $f(E,L_z)$ models fit the
observed VP shapes inferred from ground-based data to within $\sim
2$\% (in terms of deviations from a Gaussian), suggested that the M32
DF might in fact be close to the form $f(E,L_z)$. We find here that
$f(E,L_z)$ models for M32 can also fit the new HST data, and that this
requires the presence of a nuclear dark mass, as was the case for the
ground-based data. However, the best fitting dark mass of $\Mdark =
(2.5$--$4.5) \times 10^6 \Msun$ is larger than the $\Mdark = (2$--$3)
\times 10^6 \Msun$ that best fits the $\sim 0.5''$ spatial resolution
data from the CFHT, and is even more different from the $\Mdark =
(1.5$--$2) \times 10^6 \Msun$ that best-fits the $\sim 0.9''$ spatial
resolution data from the WHT. Thus, under the assumption of an
$f(E,L_z)$ DF, the different data sets cannot be fit with the same
$\Mdark$. This indicates that the M32 DF is not of the form
$f(E,L_z)$, although it might be close to it.

To obtain a model-independent estimate of the best-fitting $\Mdark$,
and to firmly rule out models without any dark mass, it is necessary
to study more general three-integral models. We have made such models
for M32, both with and without central BHs, and for various possible
values of the average mass-to-light ratio $\Upsilon$ of the stellar
population. The models were constructed to fit all available
kinematical HST, CFHT and WHT data, and the acceptability of each
model was assessed through the $\chi^2$ of its fit to the data. The
models demonstrate explicitly for the first time that there is no
axisymmetric constant mass-to-light ratio model that can fit the
kinematical data without invoking the presence of a nuclear dark mass,
independent of the dynamical structure of M32. A nuclear dark point
mass of $\Mdark = (3.4 \pm 1.6) \times 10^6 \Msun$ is required (with
$1\sigma$ and $3\sigma$ error bars of $0.7 \times 10^6 \Msun$ and $1.6
\times 10^6 \Msun$, respectively, which includes the possible effect of
small numerical errors in the models). This mass is similar to that
quoted by most previous papers, but the confidence on the detection of
a nuclear dark mass in M32 is now much higher. Constant mass-to-light
ratio models still come very close to fitting the ground-based data,
and only the new HST data make the case for a nuclear dark mass
clear-cut.

The inclination of M32 cannot be inferred from the available surface
photometry, and is therefore a free parameter in the modeling. Ideally
one would like to construct dynamical models for all possible
inclinations (which would be very computer-intensive) and determine
the inclination that best fits the kinematical data. Here we have
taken the more modest approach of constructing models for only two
representative inclinations: $i=90^{\circ}$ (edge-on) and
$i=55^{\circ}$. The intrinsic axial ratios for these inclinations are
$q=0.73$ and $q=0.55$, respectively. The three-integral $i=55^{\circ}$
models provide a better fit than the edge-on models, which suggests
that M32 is not seen edge-on. However, the allowed range for $\Mdark$
does not depend sensitively on the assumed inclination: models with no
central dark mass are firmly ruled out for both inclinations. So even
though a more detailed study of the full inclination range for M32
would improve our knowledge of the true inclination and intrinsic
axial ratio of M32, it would probably not change significantly the
constraints on the central dark mass.

The best-fitting three-integral models are similar to $f(E,L_z)$
models in that they have an excess of azimuthal motion. This is why
they have been so successful in fitting ground-based data, including
available VP shape parameters, and it confirms that they provide a
useful low-order approximation to the dynamical structure of
M32. However, $f(E,L_z)$ models do have $\sigma_r \equiv
\sigma_{\theta}$. This does not reproduce the inequality between
$\sigma_r$ and $\sigma_{\theta}$, nor the modest tilt of the velocity
ellipsoid indicated by the small $\langle v_r v_\theta \rangle$ term,
seen in the best-fitting three-integral models. This is why $f(E,L_z)$
models cannot successfully explain {\it all} observed features of the
kinematical data.

To constrain the size of the dark object in M32 we have constructed
$f(E,L_z)$ models with an extended dark nuclear object. These show
that the HST data put an upper limit of $0.08'' = 0.26
\pc$ on the half-mass radius of the nuclear dark object, implying a
central density exceeding $1 \times 10^8 \Msun \pc^{-3}$. This limit
on the density of dark material in M32 essentially rules out nuclear
clusters of planets, brown dwarfs, white dwarfs, neutron stars, or
smaller mass BHs (van der Marel 1997a). The absence of color gradients
in the central arcsec of M32 implies that the nuclear mass
concentration can also not be attributed to a stellar population
gradient. A single massive nuclear BH therefore provides the most
plausible interpretation of the data.

\subsection{Dynamical stability}

Axisymmetric dynamical models with a nuclear BH provide an excellent
fit to all available kinematical data for M32. However, to be
physically meaningful, the models must also be dynamically stable. In
van der Marel, Sigurdsson \& Hernquist (1997c) we presented N-body
simulations of the $f(E,L_z)$ models for M32. The models were found to
be completely stable, both for $i=90^{\circ}$ and for
$i=55^{\circ}$. This shows that dynamical stability is not a problem
for the models, and that the inclination of M32 cannot be meaningfully
constrained through stability arguments. We have not evolved N-body
models for the best-fitting three-integral models, but we expect these
models to be stable as well, given their similarity to $f(E,L_z)$
models.

\subsection{Dynamical relaxation}

The two-body relaxation time in M32 can be estimated as in, e.g.,
Binney \& Tremaine (1987; eq.~[8-71]). Using the relevant quantities
for our best-fitting dynamical model, we find for solar mass stars in
the central cusp ($r \lta 0.5''$) that $t_{\rm relax} \approx 3 \times
10^9 \> (r/0.1'')^{-0.065} \yr$. The time scale for `resonant
relaxation' (Rauch \& Tremaine 1996) is of the same order. The central
cusp must therefore be evolving secularly over a Hubble time. However,
the diffusion of stars in phase space is slow enough that one may
assume the evolution to be through a sequence of quasi-equilibrium
models. This justifies our approach of modeling M32 as a collisionless
equilibrium system. Studies of the secular evolution of the M32 cusp
will be interesting, but will not change the need for a nuclear dark
object. In fact, the process of dynamical relaxation supports the
presence of a dark object: without a dark object the relaxation would
proceed at a much more rapid rate that is difficult to reconcile with
observations (Lauer \etal 1992).

\subsection{Triaxiality}
\label{ss:triaxiality}

One remaining uncertainty in our dynamical modeling is the possibility
of triaxiality. After the step from spherical models to axisymmetric
models, triaxial models are the obvious next step. However, there are
several reasons to believe that for M32 this additional step will be
less important. First, M32 is known not to be spherical, but there is
no reason why it cannot be axisymmetric. There is no significant
isophote twisting in M32, and no minor axis rotation. This does not
mean that M32 cannot be triaxial (we might be observing it from one of
the principal planes), but it also does not mean that M32 needs to be
triaxial. Second, spherical constant mass-to-light ratio models
(without a nuclear dark mass) for ground-based M32 data failed to fit
by only a few $\kms$, and it was quite conceivable that axisymmetry
could fix this (which it did, cf.~Figure~\ref{f:DRfit}
below). However, axisymmetric constant mass-to-light ratio models for
the new HST data fail to fit the nuclear velocity dispersion by $>50$
km/s, and this cannot likely be fixed through triaxiality. Third,
theoretical arguments suggest that strongly triaxial models with
density cusps as steep as in M32 may not be stable, owing to the fact
that regular box-orbits are replaced by boxlets and irregular orbits
that may not be able to sustain a triaxial shape (Binney \& Gerhard
1985; Merritt \& Fridman 1996; Merritt \& Valluri 1996; see also the
review by de Zeeuw 1996). Rapidly-rotating low-luminosity elliptical
galaxies like M32 always have steep power-law cusps (Faber \etal
1997), and may therefore be axisymmetric as a class (de Zeeuw \&
Carollo 1996). This is consistent with statistical studies of their
intrinsic shapes (e.g., Merritt \& Tremblay 1996). So, apart from the
fact that triaxiality is unlikely to remove the need for a central
dark object in M32, it may even be so that M32 cannot be significantly
triaxial.

\subsection{Adiabatic black hole growth}

The growth of a black hole into a stellar system is adiabatic if it
occurs over a time scale that is `long' (see Sigurdsson, Quinlan,
\& Hernquist 1995 for a quantitative discussion) compared to the typical
orbital period of the stars. For the case of M32, the black hole
formation can be considered adiabatic if it took at least $10^6 \yr$.
Young (1980) studied the adiabatic growth of BHs in spherical
isothermal models with central density $\rho_0$ and core radius $r_0$.
The BH growth leaves the mass density at large radii unchanged, but
induces a central cusp $\rho \propto r^{-1.5}$ for $r \rightarrow
0$. The form of the density profile at intermediate radii is
determined by the dimensionless parameter ${\overline{M}}_{\bullet} \equiv
\Mdark / [{4\over3} \pi \rho_0 r_0^3]$, which measures the ratio of
the BH mass to the initial core mass. Lauer \etal (1992) showed that
the shape of the M32 brightness profile measured with HST can be well
fit with ${\overline{M}}_{\bullet} = 0.33 \pm 0.11$. The radial and density
normalization implied by the data are then $r_0 = 3.0 \pc$ and $\rho_0
= (4.2 \times 10^4) \Upsilon \Msun \pc^{-3}$. This {\it photometric}
model therefore implies that $\Mdark/\Upsilon = (1.6 \pm 0.5) \times
10^6 \Msun$. Although this result depends somewhat on the assumed
isothermality of the initial distribution (Quinlan, Hernquist
\& Sigurdsson 1995), it is quite remarkable that our best-fitting 
{\it dynamical} models have exactly $\Mdark/\Upsilon = 1.6 \times 10^6
\Msun$, for both inclinations that we studied. The M32 data are
therefore fully consistent with the presence of a BH that grew
adiabatically into a pre-existing core. This is similar to the
situation for M87 (cf.~Young \etal 1978; Harms \etal 1994).

Lee \& Goodman (1989) extended Young's calculations to the case of
rotating models. For the value of ${\overline{M}}_{\bullet}$ implied by the
photometry, their models predict a profile of $\langle v_\phi \rangle
\, / \, \langle v^2 \rangle^{1/2}$ that is approximately flat with radius
(with amplitude fixed by the axial ratio of the system). However, this
result depends very sensitively on the assumed rotation law of the
initial model. The radial variations in $\langle v_\phi \rangle \, /
\, \langle v^2 \rangle^{1/2}$ seen in our best-fitting models 
(Figure~\ref{f:vsig}) are probably equally consistent with the adiabatic
growth hypothesis.

\subsection{Tidal disruption of stars}

A star of mass $m_{\star}$ and radius $r_{\star}$ on a circular orbit
of radius $r$ will be tidally disrupted if $r \lta r_{\rm t} \equiv (2
\Mdark / m_{\star})^{1/3} r_{\star}$ (e.g., Binney \& Petit 1988).
Thus, disruption of a solar type star by the BH in M32 will occur
inside $r_{\rm t} = 4.2 \times 10^{-6} \pc = 1.2 \times 10^{-6}$
arcsec. A disruption event will be highly luminous, but is not
predicted to occur more often than once every $10^4 \yr$ (Rees
1988). The minimum pericenter distance for a star with given $(E,L_z)$
in a Kepler potential is $r_{\rm p,min} = \Rc(E) \>
(1-\sqrt{1-\eta^2})$, where as before, $\Rc$ is the radius of the
circular orbit at the given energy and $\eta \equiv L_z / L_{\rm
max}(E)$. The kinematical data for M32 only meaningfully constrain the
DF for energies with $\Rc(E) \gta 0.1''$. For $\Rc(E) = 0.1''$, only
stars with $\vert \eta \vert < 5 \times 10^{-3}$ have $r_{\rm p,min} <
r_t$. The data do not constrain variations in the DF over such a small
range in $\eta$, and our dynamical models therefore cannot address the
existence and properties of the so-called `loss cone' (Frank \& Rees
1976; Lightman \& Shapiro 1977). For the $\eta$-grid that we have
employed, all solar type stars on orbits with $\Rc(E) > 0.1''$ have
$r_{\rm p,min} > 2 \times 10^2 \> r_t$. Even giants with $r_{\star}
\approx 10^2 r_{\odot}$ have $r_{\rm p,min} > r_t$. This justifies our
neglect of tidal disruption in the orbit calculations.

\subsection{Accretion onto the black hole}

An interesting question is why BHs in quiescent galaxies aren't more
luminous (e.g., Kormendy \& Richstone 1995). For M32, the total X-ray
luminosity is $L_{\rm X} \approx 10^{38} \ergs$ (Eskridge, White \&
Davis 1996), the far infrared luminosity is $L_{\rm FIR} < 3
\times 10^{36} \ergs$ (Knapp \etal 1989), and for the $6 \cm$ radio
emission $\nu L_{\nu} < 3 \times 10^{33} \ergs$ (Roberts \etal
1991). Part or all of the observed X-ray emission may be due to
low-mass X-ray binaries, so the total luminosity due to accretion onto
the BH in M32 is $L_{\rm acc} < 10^{38} \ergs$. By contrast, the
Eddington luminosity of the BH is $L_{\rm Edd} = 4.3 \times 10^{44}
\ergs$. For a canonical mass-loss rate of $1.5 \Msun 
(10^{11} \Lsun)^{-1} \yr^{-1}$ (Faber \& Gallagher 1976), the stars
that are bound to the BH in M32 shed $1 \times 10^{-4} \Msun
\yr^{-1}$ of gas as a result of normal stellar evolution. If a
fraction $f$ of this gas is steadily accreted with efficiency
$\epsilon$, it produces a luminosity $L_{\rm acc} = \epsilon f \> (6.7
\times 10^{42}) \ergs$. Thus either the accretion fraction $f$ or the
accretion efficiency $\epsilon$ must be very small in M32. Thin disk
accretion with $\epsilon \approx 0.1$ requires $f < 1.5
\times 10^{-4}$, which is possible (the accretion fraction is
difficult to predict theoretically, because it depends on the
hydrodynamics of the stellar winds that shed the gas), but may be
implausibly low. Instead, it appears more likely that $\epsilon$ is
small, since there is a family of `advection dominated' accretion
solutions that naturally predict such low efficiencies. Models of this
type successfully explain the `micro-activity' of the BH (Sgr
A$^{^\star}$) in our own Galaxy (Narayan, Yi \& Mahadevan 1995). In a
typical accretion model of this type (Narayan \& Yi 1995, their
Fig.~11), $f \lta 0.16$ suffices to explain the upper bound on $L_{\rm
acc}$ for M32.

\subsection{Forthcoming observations}

Future observations of M32 will include spectra with the new long-slit
HST spectrograph STIS. These will provide significantly better sky
coverage than our FOS data, but the spatial resolution will be
similar. The high-resolution HST data can be complemented with that
from fully two-dimensional ground-based spectrographs, such as OASIS
on the CFHT and SAURON on the WHT. These combined data will yield
improved constraints on the BH mass, on the orbital structure and
inclination of M32, and on possible deviations from axisymmetry.


\acknowledgments

We dedicate this paper, and its companions R97 and C97, to the memory
of Martin Schwarzschild, who pioneered the modeling technique employed
here. Martin's sense of purpose, his exceptional clarity of thinking,
his transparent personal integrity and, most of all, his genuine warm
interest and support, remain a great source of inspiration to us.
Support for this work was provided by NASA through grant number
\#GO-05847.01-94A, and through a Hubble Fellowship \#HF-1065.01-94A
awarded to RPvdM, both from the Space Telescope Science Institute
which is operated by the Association of Universities for Research in
Astronomy, Incorporated, under NASA contract NAS5-26555. NC
acknowledges financial support from the Swiss government (Etat du
Valais) and NUFFIC, and the hospitality of Steward Observatory, the
MPA Garching and Geneva Observatory. TdZ is grateful for the generous
hospitality of the Institute for Advanced Study. Both he and NC
received financial support from the Leids Kerkhoven Bosscha Fonds.


\clearpage
 
\appendix

\section{$\chi^2$ topology for orbit-superposition models}

In this Appendix we discuss the topology of the
$\chi^2(\Upsilon,\Mdark)$ contours for the edge-on orbit-superposition
models. The top panels of Figure~{\ref{f:subdatachi}} show the
$\chi^2$ contours when only (subsets of) the ground-based WHT data are
included in the fit. These panels can be compared to
Figure~{\ref{f:subdatachi}d}, which shows the contours for the case in
which all WHT, CFHT and HST data are included.

Figure~{\ref{f:subdatachi}a} shows the $\chi^2$ contours when only the
major axis $V$ and $\sigma$ WHT measurements are fit. Binney \& Mamon
(1982) showed that a large range of gravitational potentials can fit
any given observed velocity dispersion profile. The valley seen in the
$\chi^2$ contours is a consequence of this: it outlines a
one-parameter family of models that can fit the data with different
velocity dispersion anisotropy. For a non-rotating spherical system,
only models that require negative second velocity moments are ruled
out. For a rotating system like M32, the observed rotation rate sets
additional limits on the allowed radial anisotropy. For the case of
the major axis $V$ and $\sigma$ WHT measurements, a no-BH model is
just marginally acceptable at $99.73$ per cent confidence,
cf.~Figure~{\ref{f:subdatachi}a}. For the lower-spatial resolution
major axis $V$ and $\sigma$ measurements of Dressler \& Richstone
(1988) such a model is entirely acceptable.  Figure~{\ref{f:DRfit}}
compares the predictions of the best-fit axisymmetric
orbit-superposition model without a BH to their data.  Richstone,
Bower \& Dressler (1990) concluded that these data could not be fit by
any {\it spherical} model without a BH. This is because spherical
models allow less rotation, and therefore failed to fit the observed
rotation velocities. This underscores the importance of making
axisymmetric models for flattened galaxies like M32.

VP shape measurements provide independent constraints on the velocity
dispersion anisotropy. Figure~{\ref{f:subdatachi}b} shows the $\chi^2$
contours for edge-on orbit-superposition models when not only the WHT
major axis $V$ and $\sigma$ measurements are fit, but also the major
axis VP shape measurements. With the inclusion of the VP shapes,
models without a BH are ruled out. Figure~{\ref{f:subdatachi}c} shows
the $\chi^2$ contours when also the WHT measurements along other
position angles are included, which contracts the allowed $\Mdark$
range to $(1.1$--$5.1) \times 10^6 \Msun$ at the formal $99.73$ per
cent confidence level. The WHT data by themselves therefore rule out
axisymmetric models without a BH.  However, the models without a BH
still come very close to fitting the data, and, e.g., fail to fit the
central velocity dispersion by only 1--2$\kms$
(cf.~Figure~\ref{f:datafitsAtoH}). So one cannot make a particularly
strong claim for a BH on the basis of the WHT data alone, because it
is conceivable that the fit could be improved with, e.g., only a minor
amount of triaxiality. The same holds for the CFHT data, but the new
HST data do make the case for a BH in M32 clear-cut.

The contours for the case in which all the available WHT, CFHT and HST
kinematical data are included in the fit
(Figure~{\ref{f:subdatachi}d}) show one global $\chi^2$ minimum, and a
second local minimum. The presence of a global minimum does not
necessarily imply that the combined data constrain a single best-fit
potential. It might be that there is a small range of potentials that
all fit equally well, but that such a range of constant $\chi^2$ would
not be evident due to the finite numerical accuracy of our
technique. In Figure~{\ref{f:subdatachi}e} we show explicitly how the
topology of the $\chi^2$ contours might have been influenced by the
possibility of small numerical errors in our models. It was obtained
from Figure~{\ref{f:subdatachi}d} by recalculating the $\chi^2$
contours after adding a random error $\Delta V,\sigma \in [-2,2]$ km/s
and $\Delta h_i \in [-0.01, 0.01]$ (cf.~Section~\ref{s:technique}) to
the prediction for each data point, for each $(\Upsilon,\Mdark)$
combination. The results show that numerical errors can indeed
influence the $\chi^2$ contours near the $\chi^2$ minimum. Thus, the
second minimum in Figure~{\ref{f:subdatachi}d} might be the result of
numerical inaccuracies in our technique. However, the numerical errors
are small enough that they only have a negligible effect on the
overall $\chi^2$ topology. In particular, models without a dark mass
remain firmly ruled out.

To assess the possible effect of numerical errors on the confidence
bands for $\Mdark$, we constructed 100 figures like
Figure~{\ref{f:subdatachi}e} using different random realizations.  For
each we determined the position of the $\chi^2$ minimum, and the
minimum and maximum $\Mdark$ for which there is an $\Upsilon$ such
that the model with $(\Mdark,\Upsilon)$ falls within the $99.73$ per
cent confidence region. The results are plotted in
Figure~{\ref{f:subdatachi}f}. All allowed $\Mdark$ values fall in the
range $\Mdark = (1.8$--$5.0) \times 10^6 \Msun$. Thus, $\Mdark = (3.4
\pm 1.6) \times 10^6 \Msun$ at $99.73$ per cent confidence. Similar 
experiments show that $\Mdark = (3.4 \pm 0.7) \times 10^6 \Msun$ at
$68.3$ per cent confidence. Experiments for $i=55^{\circ}$ produced
similar results, and mass ranges that were either the same or slightly
smaller. Thus, we conclude that the $1\sigma$ and $3\sigma$ errors on
the estimated $\Mdark = 3.4 \times 10^6 \Msun$, are $0.7$ and $1.6
\times 10^6 \Msun$, respectively.

\clearpage


\clearpage



\def\figcapone{Data points are observations of the M32 major axis
V-band surface brightness from Lauer \etal (1992), Peletier (1993) and
Kent (1987). The R-band data from Kent and Peletier were transformed
to the V-band by assuming a constant V$-$R color.  The differences
between the data sets at large radii are due to uncertainties in the
sky subtraction. Measurements are not plotted at radii where the PSF
introduces large uncertainties ($\sim 0.1''$ for the Lauer \etal HST
data, $\sim 2''$ for the ground-based data). The solid curve is the
brightness profile for the axisymmetric luminous density model used in
the present paper. The dashed curve is the profile for the model used
by van der Marel \etal (1994b) and Qian \etal (1995).\label{f:surfbr}}


\def\figcaptwo{Data points show the rotation velocities $V$ and velocity
dispersions $\sigma$ derived from the HST/FOS data of Paper~I, and
from the ground based WHT and CFHT data of van der Marel \etal (1994a)
and Bender, Kormendy \& Dehnen (1996). Errors for the ground-based
data are $\sim 1 \kms$ for the WHT data and $\sim 6 \kms$ for the CFHT
data, but are not plotted for clarity. The abscissa $r$ is the major
axis distance.  Curves show the predictions for the HST setup of
edge-on $f(E,L_z)$ models with no nuclear dark mass (dashed curves)
and with nuclear point masses (BHs) of 1, 2, 3, 4, 5 and $6 \times
10^6 \Msun$ (solid curves). Models with $\Mdark \approx (3 \pm 1)
\times 10^6 \Msun$ best reproduce the trend in the observed
dispersions.\label{f:HSTBHpred}}


\def\figcapthree{The relative RMS velocity dispersion residual
$\chi_{\sigma,{\rm RMS}}$ for edge-on $f(E,L_z)$ models with nuclear
point masses, as function of $\Mdark$. Results are shown for the HST
dispersion measurements of Paper~I, and for the ground-based WHT and
CFHT measurements. The $\chi_{\sigma,{\rm RMS}}$ for the best fit to
the WHT data is larger than for the CFHT data, because the WHT data
have much smaller errors (such that differences between the
predictions and the data are statistically more significant). The
symbols in the box mark the position of the best-fitting $\Mdark$ for
each data set and its formal $1\sigma$ error bar (the latter is
determined not only by the curvature at the $\chi_{\sigma,{\rm RMS}}$
minimum, but also by the number of data points, which is different for
each data set). Models with $\Mdark \approx 3 \times 10^6 \Msun$ best
fit the HST and the CFHT data. The WHT data are not well fitted by
this $\Mdark$. The fact that data of different spatial resolution
cannot be fit with the same $\Mdark$ implies that M32 does not have a
DF of the form $f=f(E,L_z)$.\label{f:chisqBH}}


\def\figcapfour{Horizontally hatched regions show the average velocity
dispersion and its $1\sigma$ error for those HST, CFHT and WHT
observations with the aperture center within $0.1''$ from the M32
nucleus. Curves show the predictions for edge-on $f(E,L_z)$ models
with nuclear point masses, as function of $\Mdark$. Models with no
nuclear dark mass are ruled out under the $f(E,L_z)$
hypothesis. Models with nuclear point masses reproduce the increase in
the observed nuclear velocity dispersion with increasing spatial
resolution. The horizontal bars in the boxed region indicate for each
data set the range of $\Mdark$ values that predict a nuclear
dispersion within the observed hatched region. As in
Figure~{\reffchisqBH}, the data of different spatial resolution cannot
be fit with one single value of $\Mdark$.\label{f:sigcen}}


\def\figcapfive{Example of the meridional plane at a fixed energy $E$.
The axes are in units of $\Rc(E)$, the radius of the circular orbit at
the given energy. The library in our orbit-superposition models uses
an open grid of 7 values of $\vert \eta \vert \in [0,1]$, where $\eta
\equiv L_z/L_{\rm max}(E)$; orbits for $L_z>0$ and $L_z<0$ are 
identical, except for a reversal of the three-dimensional velocity
vector at each phase-point. The oval curves are the zero-velocity
curves (ZVCs) for each $\vert \eta \vert$. Positions on the ZVC can be
parametrized using the angle $w$. Because every orbit with $L_z
\not=0$ touches the ZVC (Ollongren 1962), all orbits at a given
$(E,L_z)$ can be sampled by starting stars with $v_R=v_z=0$ from the
ZVC (but $v_{\phi} \not= 0$ for $L_z \not= 0$). The `thin tube' orbit
(heavy solid curve, for the case of the highlighted ZVC) is the only
orbit that touches the ZVC at only a single value of $\vert w \vert$,
referred to as $w_{\rm th}$. All other orbits touch the ZVC at at
least two values of $\vert w \vert$, one smaller than $w_{\rm th}$ and
one larger than $w_{\rm th}$ (see C97 for examples of actual orbits).
Orbits with $w \in [-\pi,0]$ follow trivially from those with $w \in
[0,\pi]$ upon multiplication of $(z,v_z)$ by $-1$, at each phase-point
along the orbit. It is therefore sufficient to consider only orbits
with $w \in [0,w_{\rm th}]$. We sample this range using an open grid
of 7 values (indicated by the dots, for the case of the highlighted
ZVC). The figure shows the meridional plane at the energy for which
$\Rc(E) = 0.25''$, in the edge-on model with $\Mdark = 3 \times 10^6
\Msun$ and $\Upsilon = 2.51$. The ZVCs at other energies and in other
models differ in the details, but are topologically
similar.\label{f:zvcplot}}


\def\figcapsix{Contour plots of the $\chi^2$ that measures the quality
of the fit to the combined HST, CHFT and WHT data, for
orbit-superposition models with $i=90^{\circ}$ and $i=55^{\circ}$. The
model parameters along the abscissa and ordinate are the BH mass
$\Mdark$ and the (V-band) mass-to-light ratio $\Upsilon$,
respectively. The dots indicate models that were calculated, the
contours were obtained through spline interpolation (the first three
contours define the formal $68.3$, $95.4$ and (heavy contours) $99.73$
per cent confidence regions; subsequent contours are characterized by
a factor two increase in $\Delta \chi^2$). The bottom right corner in
each panel is a region for which no models were calculated. The
labeled positions define models that are discussed in detail in the
text and in subsequent figures. Models~C and~G provide the overall
best fits.\label{f:contour}}


\def\figcapseven{Predictions of the best-fitting orbit-superposition models
for $i=90^{\circ}$ and $i=55^{\circ}$ (labeled~C and~G in
Figure~{\reffcontour}), compared to the kinematical HST, CFHT and WHT
data. The models have nuclear BHs of $3.4 \times 10^6 \Msun$ and $3.2
\times 10^6 \Msun$, respectively. The HST data have the highest
spatial resolution, and were taken with a set of apertures aligned
along the major axis. The ground-based data are long-slit
measurements. For the CFHT observations two independent sets of major
axis data are available with a similar setup. For the WHT observations
data are available with the slit along the major axis, minor axis, two
intermediate axes (major $\pm 45^{\circ}$), and an axis parallel to
the major axis but $4''$ offset from it. Shown from top to bottom are:
rotation velocities, velocity dispersions, and the Gauss-Hermite
moments (when available) that measure deviations of the line-of-sight
VP shapes from a Gaussian. The data points are arranged equidistantly
along the abscissa. The corresponding distance from the nucleus in
arcsec is illustrated schematically in the bottom panel. The WHT data
were analyzed by averaging spectra at positive and negative radii, so
for these data only positive radii are shown. The model fits to the
data are excellent.\label{f:bestfits}}

\def\figcapsevenB{(continued).}


\def\figcapeight{Predicted rotation velocities and velocity
dispersions for the orbit-superposition models~A--D (top panel)
and~E--H (bottom panel) defined in Figure~{\reffcontour}. Models~C\&G
are the overall best fits, models~A\&E are the best fits without a
central dark mass, and models~B\&F and~D\&H are (approximately) the
best-fitting models for $\Mdark=1.9$ and $5.4 \times 10^6 \Msun$,
respectively. The data are as in Figure~{\reffbestfits}. The models
without a BH manage to fit the WHT data reasonably well, but are
firmly ruled out by the HST data.\label{f:datafitsAtoH}}


\def\figcapnine{The top row shows the $(\eta,w)$ integral space 
(defined as in Figure~{\reffzvcplot}) for a selected set of energies,
for the best-fitting edge-on orbit-superposition model~C defined in
Figure~{\reffcontour}. Each square in each panel represents an
orbit. The (logarithmic) grey-scale shows the fraction of the mass at
the given energy that was assigned to each orbit by the NNLS
fit. Smoother solutions are obtained by adding regularization
constraints to the NNLS fit. The second and third rows show the
integral space for the same model with a modest and a large amount of
regularization, respectively. Most of the mass resides at $\eta >0$
(i.e., $L_z>0$), which is obviously required to fit the observed
rotation of M32. The bottom panels show the fits of the models to the
observed rotation velocities and velocity dispersions. The model
without regularization provides the best fit, but even for the
smoothest model the fits are still quite good.\label{f:intspace}}

\def\figcapnineB{(continued).}


\def\figcapten{Velocity moments for the orbit-superposition 
models~A--D and~E--H defined in Figure~{\reffcontour}, averaged over
spherical shells. The displayed results were obtained with a modest
amount of regularization in the NNLS fit, to obtain smoother results.
The rightmost panels show the predictions for $f(E,L_z)$ models with
the same gravitational potential as models~C\&G, obtained by solving
the Jeans equations. Only those radii are shown for which the
dynamical structure of the models is meaningfully constrained by
kinematical data. Models~A\&E have no BH and invoke as much radial
motion as possible (under the constraint that the rotation curve is
fit) to produce a peak in the observed velocity
dispersions. Nonetheless, they cannot fit the data
(cf.~Figure~{\reffdatafitsAtoH}). The models with a BH are all
dominated by azimuthal motion at most radii. Models with an $f(E,L_z)$
DF provide a useful approximation to the dynamics of M32 because they
also have this property. However, the $f(E,L_z)$ models have $\vthsqav
\equiv \vrsqav$, which is not the case for the models that best fit
the data.\label{f:moments}}


\def\figcapeleven{The quantity $\langle v_\phi \rangle \, / \,
\langle v^2 \rangle^{1/2}$ as function of radius in the equatorial plane,
for the models~C and~G defined in Figure~{\reffcontour}. Solid curves
show the predictions for the orbit superposition models.  As in
Figure~{\reffmoments}, a modest amount of regularization was used in
the NNLS fit, and only those radii are shown for which the dynamical
structure of the models is meaningfully constrained by kinematical
data. Dotted curves show the predictions for oblate isotropic rotator
models, obtained by solving the Jeans equations. The inclined, and
intrinsically flatter, model~G rotates faster than the edge-on
model~C.\label{f:vsig}}


\def\figcaptwelve{Solid curves show the velocity dispersions predicted 
for the HST setup by edge-on $f(E,L_z)$ models with an extended
nuclear dark mass. Data points are as in Figure~{\reffHSTBHpred}. The
models in the top panel have $\Mdark = 3 \times 10^6 \Msun$, those in
the bottom panel have $\Mdark = 6 \times 10^6 \Msun$. The models have
nuclear dark objects with scale radii of $\epsilon = 0$, $0.04''$,
$0.1''$, $0.25''$ and $0.5''$, as indicated. The models with the
smallest $\epsilon$ best fit the data.\label{f:HSTextpr}}


\def\figcapthirteen{Contour plots of the quantity $\chi_{\sigma}^2$ that 
measures the quality of the model fit to the HST velocity dispersion
measurements, for $f(E,L_z)$ models with an extended dark nuclear
object and an inclination of $i=90^{\circ}$ (edge-on) or
$i=55^{\circ}$, respectively. The model parameters along the abscissa
and ordinate are the nuclear dark mass $\Mdark$ and its scale radius
$\epsilon$, respectively. The contours are defined as in
Figure~{\reffcontour}. The results show that the nuclear dark mass in
M32 must be less extended than $\epsilon = 0.06''$, independent of the
inclination.\label{f:chisqsurf}}


\def\figcapfourteen{Contour plots of $\chi^2(\Mdark,\Upsilon)$ for
edge-on orbit superposition models constructed to fit: (a) the major
axis WHT $V$ and $\sigma$ measurements; (b) all major axis WHT
measurements, including VP shapes; (c) all WHT measurements, including
various position angles; (d) all WHT, CFHT and HST data (same as
Figure~\ref{f:contour}). Contours are defined as in
Figure~\ref{f:contour}. Heavy contours show the formal $99.73$\%
confidence regions. Panel~(e) shows how the contours in panel~(d) are
modified if random errors are added to the predictions for each data
point, to simulate numerical errors in the models. Panel~(f) shows for
100 simulations as in panel~(e) the position of the $\chi^2$ minimum
(solid symbols), and the lowest and highest $\Mdark$ that fall within
a $99.73$\% confidence contour (open symbols). These simulations show
that numerical errors cannot be responsible for the fact that models
with either $\Mdark < 1.8 \times 10^6 \Msun$ or $\Mdark > 5.0 \times
10^6 \Msun$ fail to fit the data at this confidence
level.\label{f:subdatachi}}


\def\figcapfifteen{Rotation velocities and velocity dispersions 
for the edge-on axisymmetric orbit superposition model without a BH
that best fits the data of Dressler \& Richstone (1988). These data
could not be fit by any spherical model without a BH (Richstone, Bower
\& Dressler 1990). It is thus important to make axisymmetric models 
for flattened galaxies like M32.\label{f:DRfit}}


\ifsubmode
\figcaption{\figcapone}
\figcaption{\figcaptwo}
\figcaption{\figcapthree}
\figcaption{\figcapfour}
\figcaption{\figcapfive}
\figcaption{\figcapsix}
\figcaption{\figcapseven}
\figcaption{\figcapeight}
\figcaption{\figcapnine}
\figcaption{\figcapten}
\figcaption{\figcapeleven}
\figcaption{\figcaptwelve}
\figcaption{\figcapthirteen}
\figcaption{\figcapfourteen}
\figcaption{\figcapfifteen}
\clearpage
\else\printfigtrue\fi

\ifprintfig


\begin{figure}
\centerline{\epsfbox{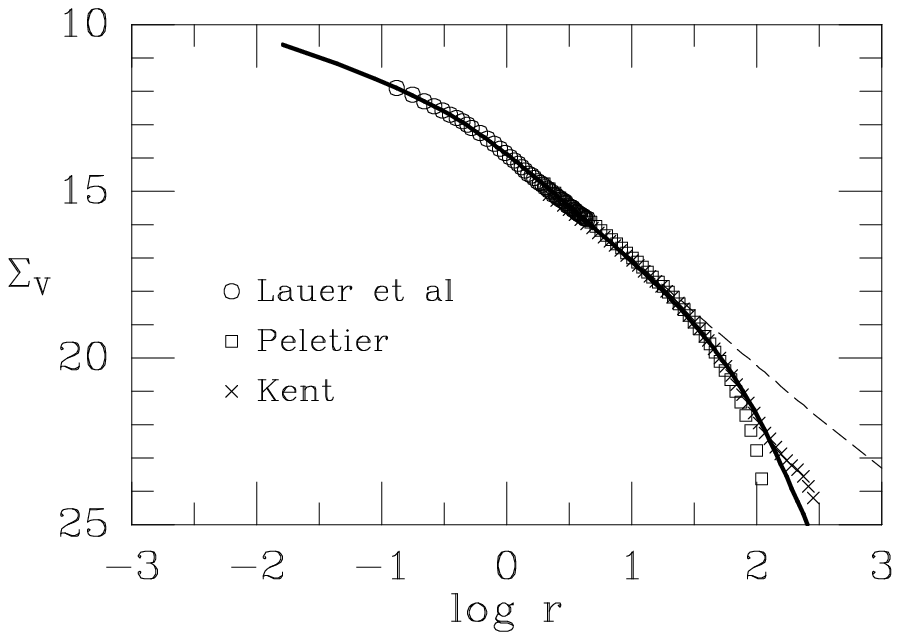}}
\ifsubmode
\vskip3.0truecm
\centerline{Figure~1}\clearpage
\else\figcaption{\figcapone}\fi
\end{figure}


\begin{figure}
\centerline{\epsfbox{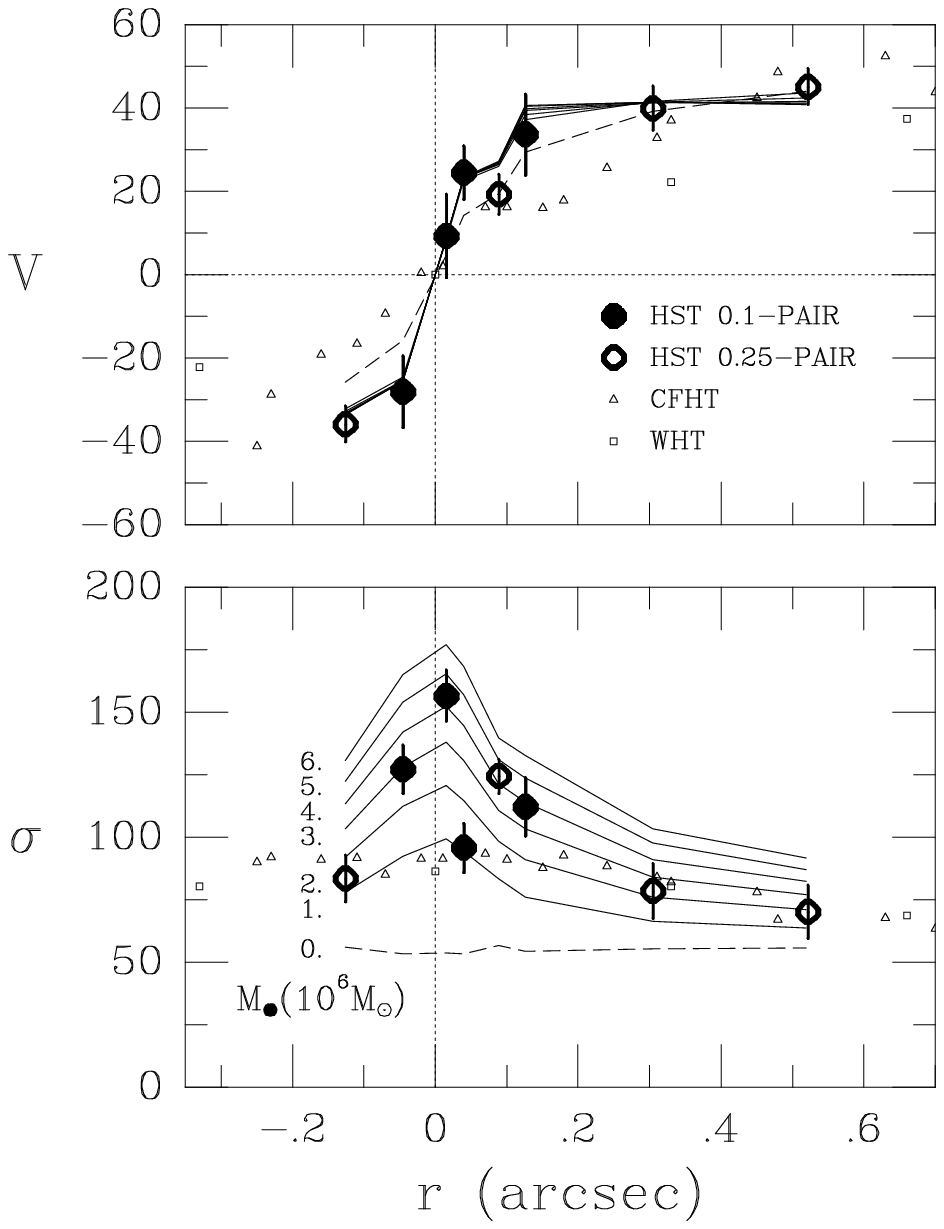}}
\ifsubmode
\vskip3.0truecm
\centerline{Figure~2}\clearpage
\else\figcaption{\figcaptwo}\fi
\end{figure}


\begin{figure}
\centerline{\epsfbox{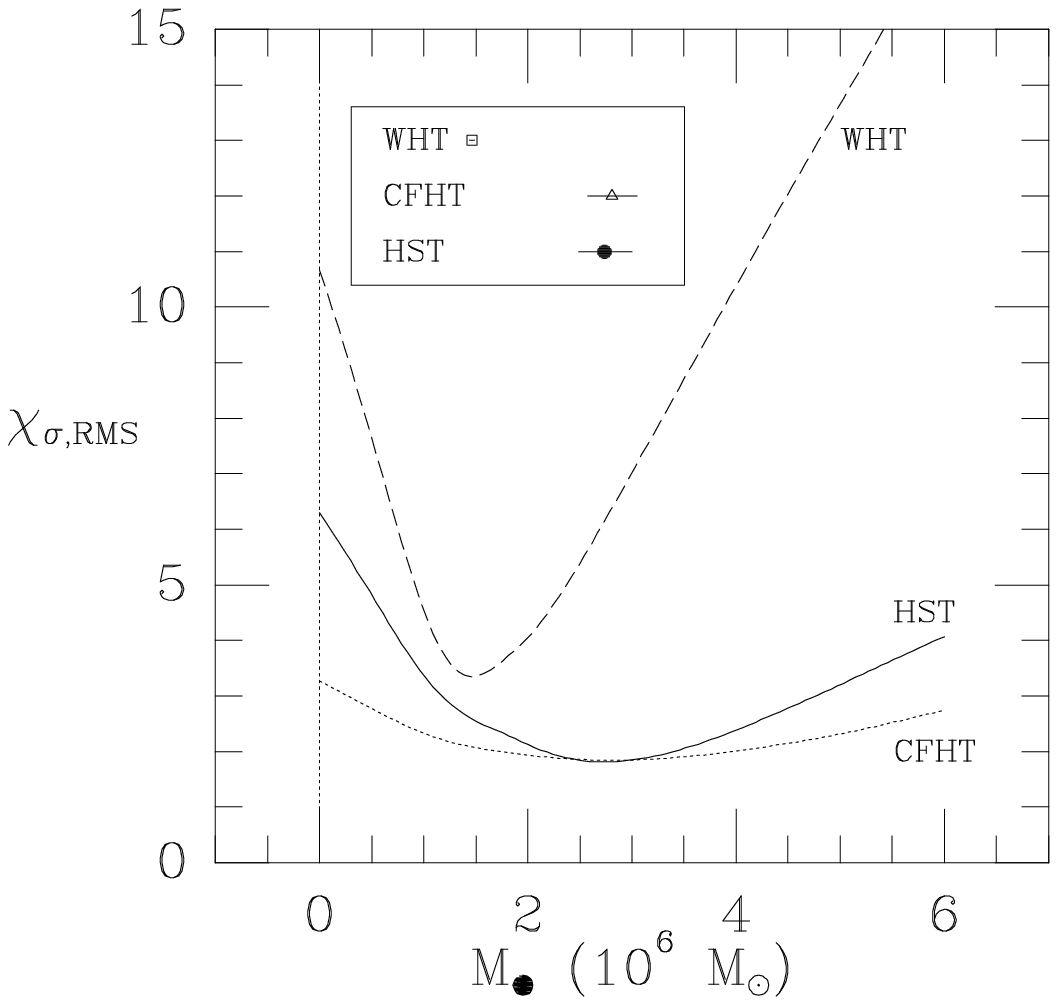}}
\ifsubmode
\vskip3.0truecm
\centerline{Figure~3}\clearpage
\else\figcaption{\figcapthree}\fi
\end{figure}


\begin{figure}
\centerline{\epsfbox{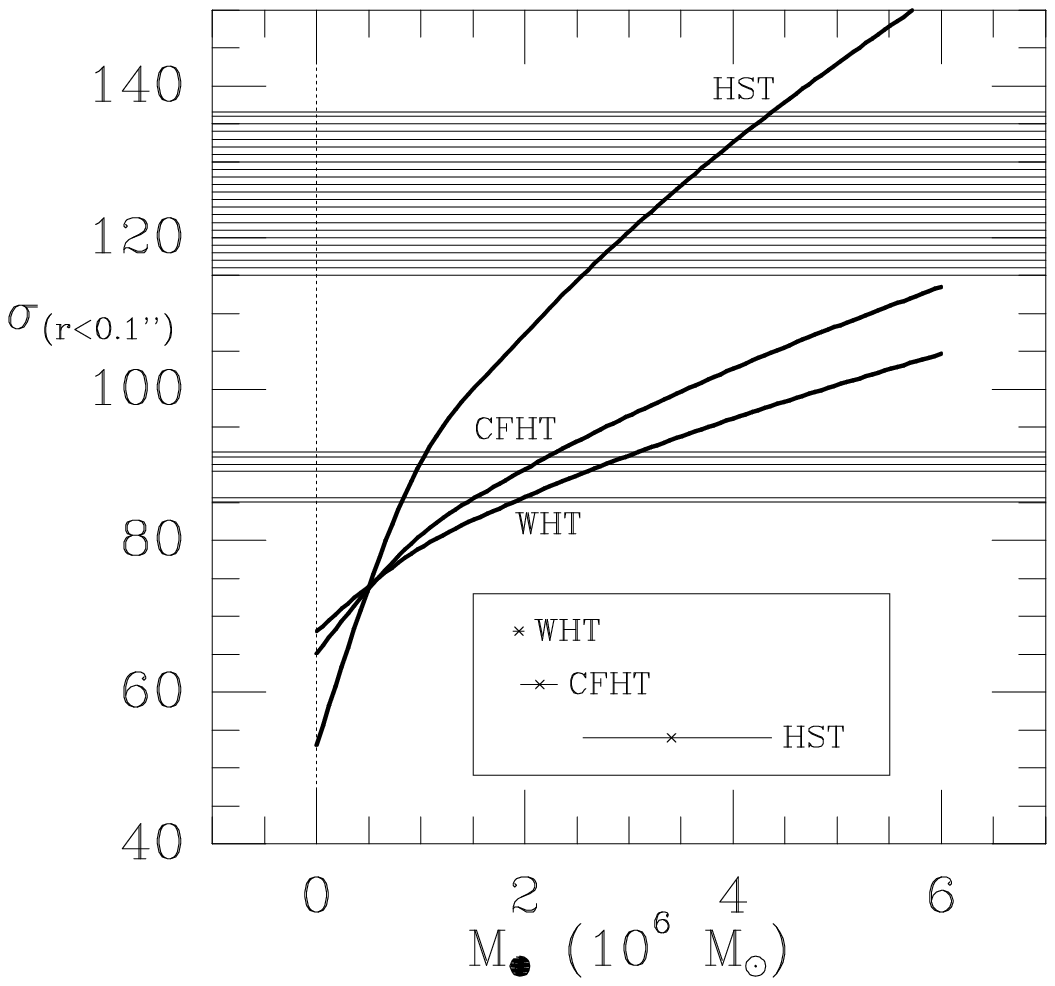}}
\ifsubmode
\vskip3.0truecm
\centerline{Figure~4}\clearpage
\else\figcaption{\figcapfour}\fi
\end{figure}


\begin{figure}
\epsfxsize=12.0truecm
\centerline{\epsfbox{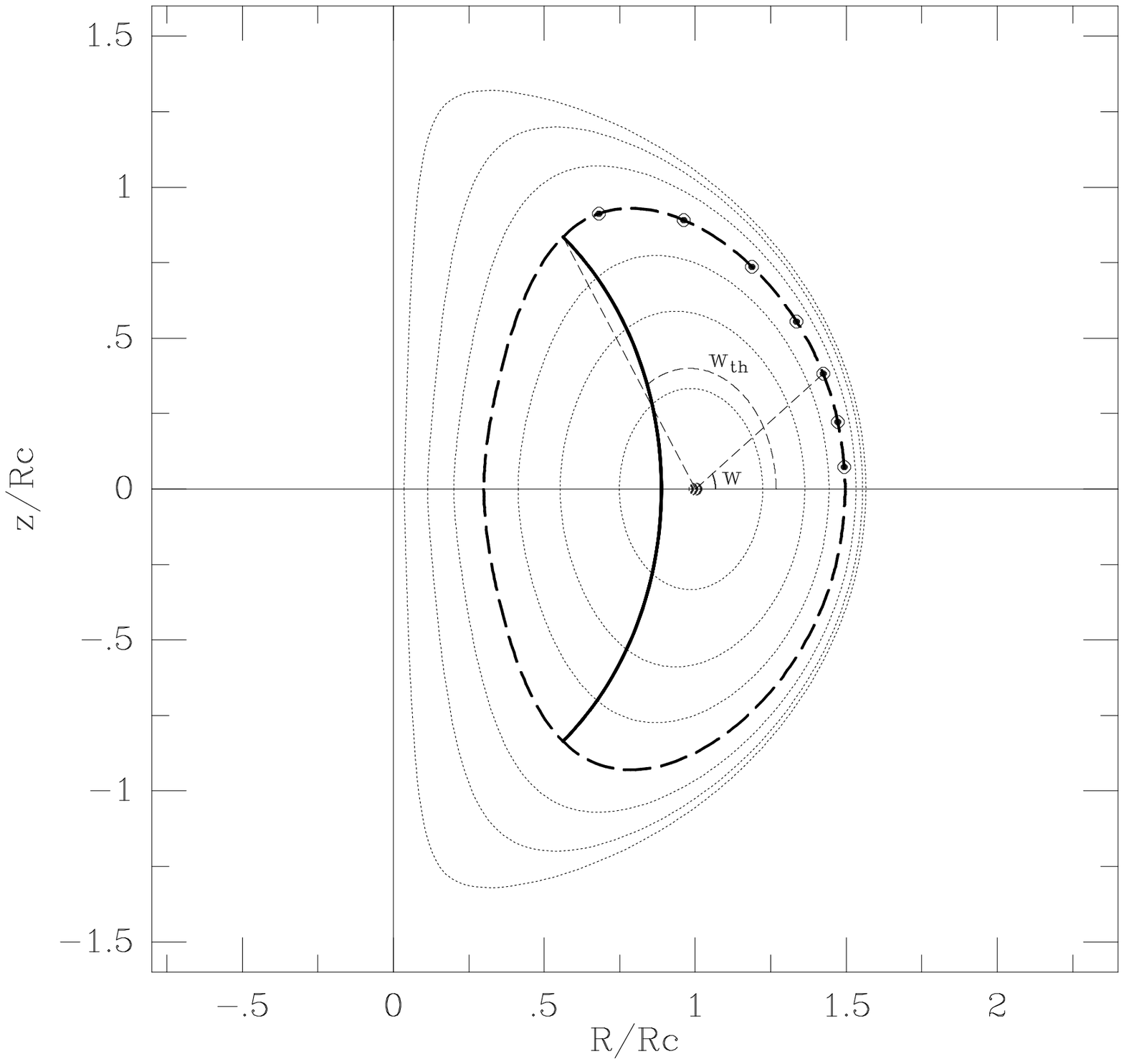}}
\ifsubmode
\vskip3.0truecm
\centerline{Figure~5}\clearpage
\else\figcaption{\figcapfive}\fi
\end{figure}


\begin{figure}
\epsfxsize=16.0truecm
\centerline{\epsfbox{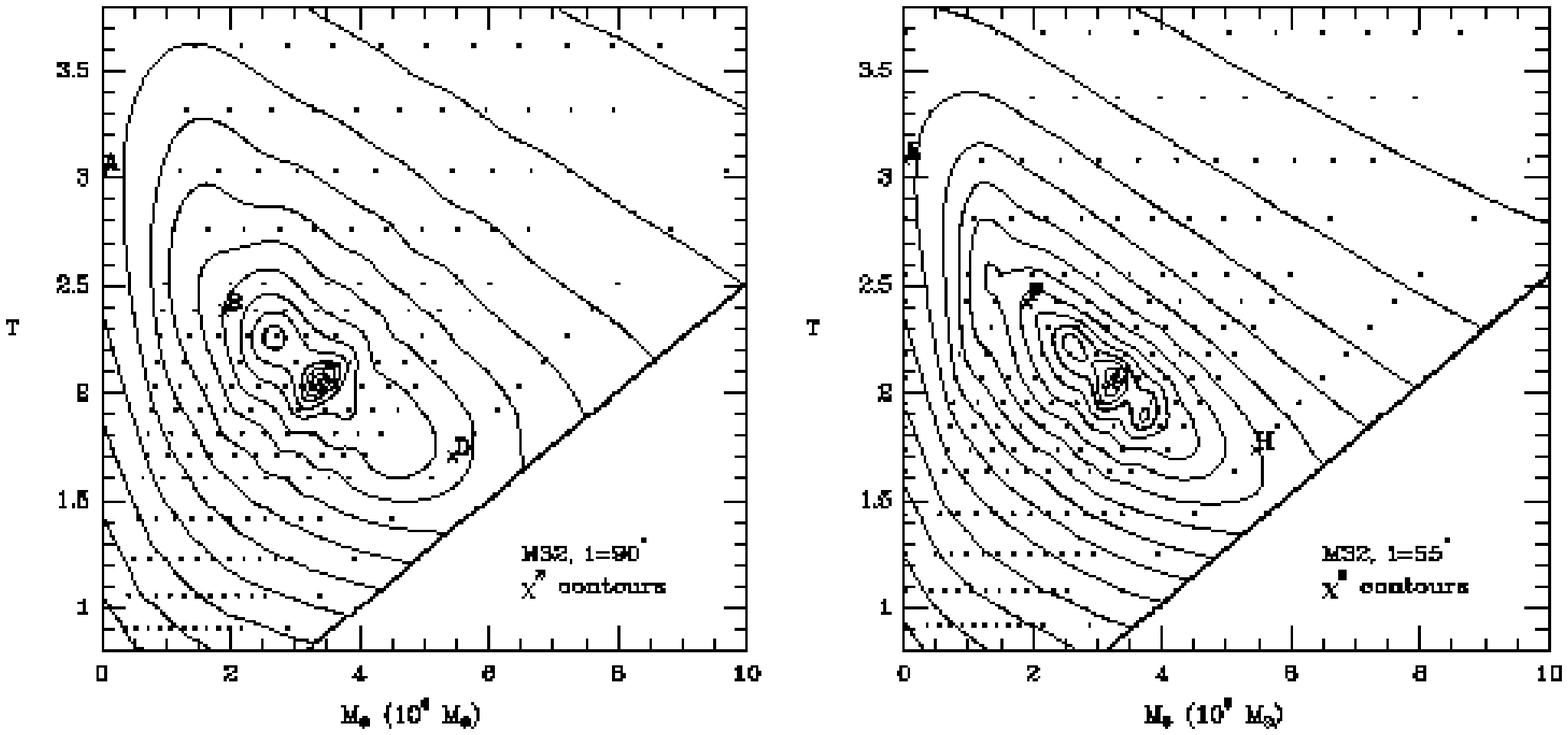}}
\ifsubmode
\vskip3.0truecm
\centerline{Figure~6}\clearpage
\else\figcaption{\figcapsix}\fi
\end{figure}


\begin{figure}
\ifsubmode
\else\figcaption{\figcapseven}\fi
\end{figure}

\begin{figure}
\figurenum{7}
\epsfxsize=15.0truecm
\centerline{\epsfbox{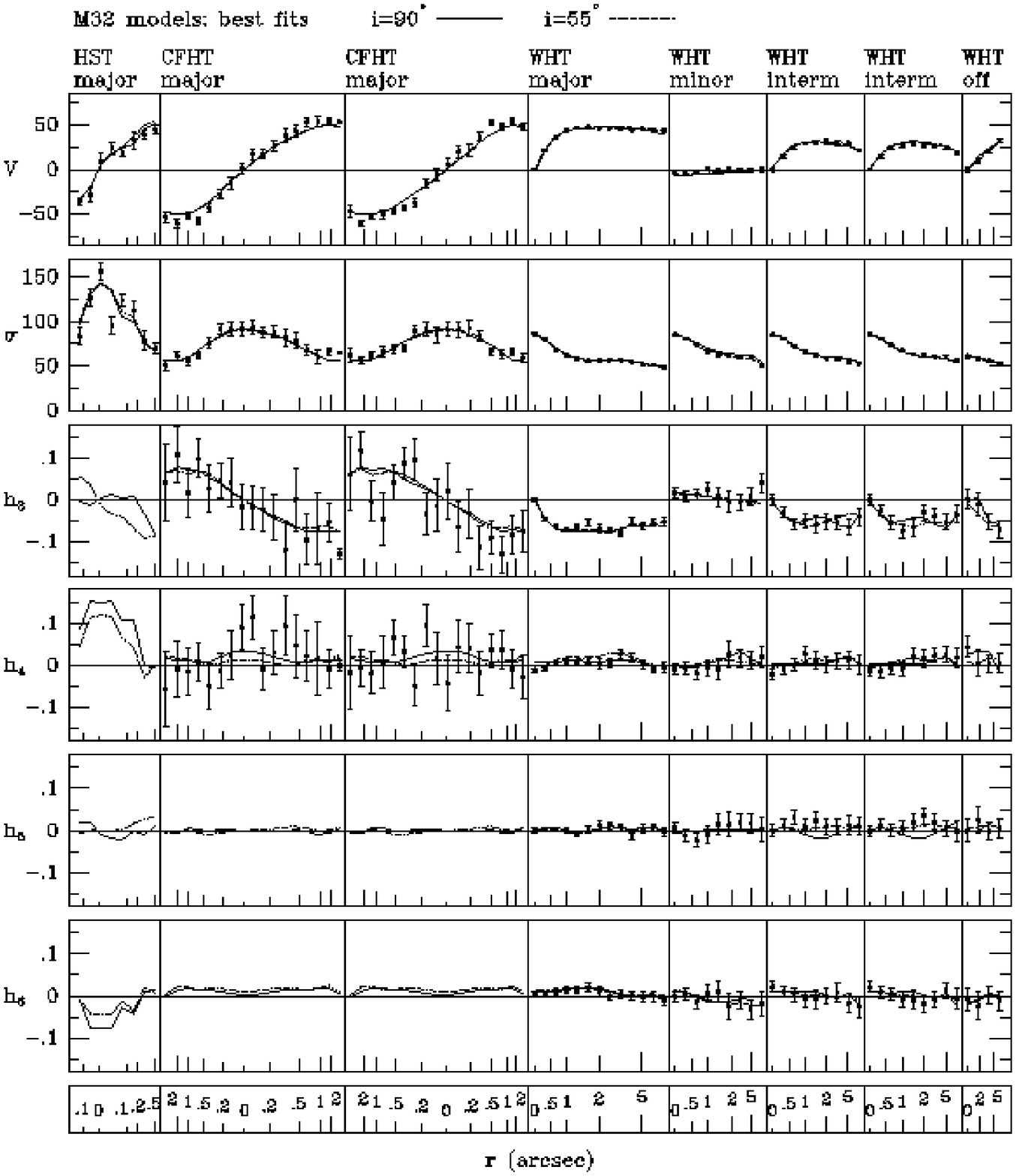}}
\ifsubmode
\vskip3.0truecm
\centerline{Figure~7}\clearpage
\else\figcaption{\figcapsevenB}\fi
\end{figure}


\begin{figure}
\epsfxsize=12.0truecm
\centerline{\epsfbox{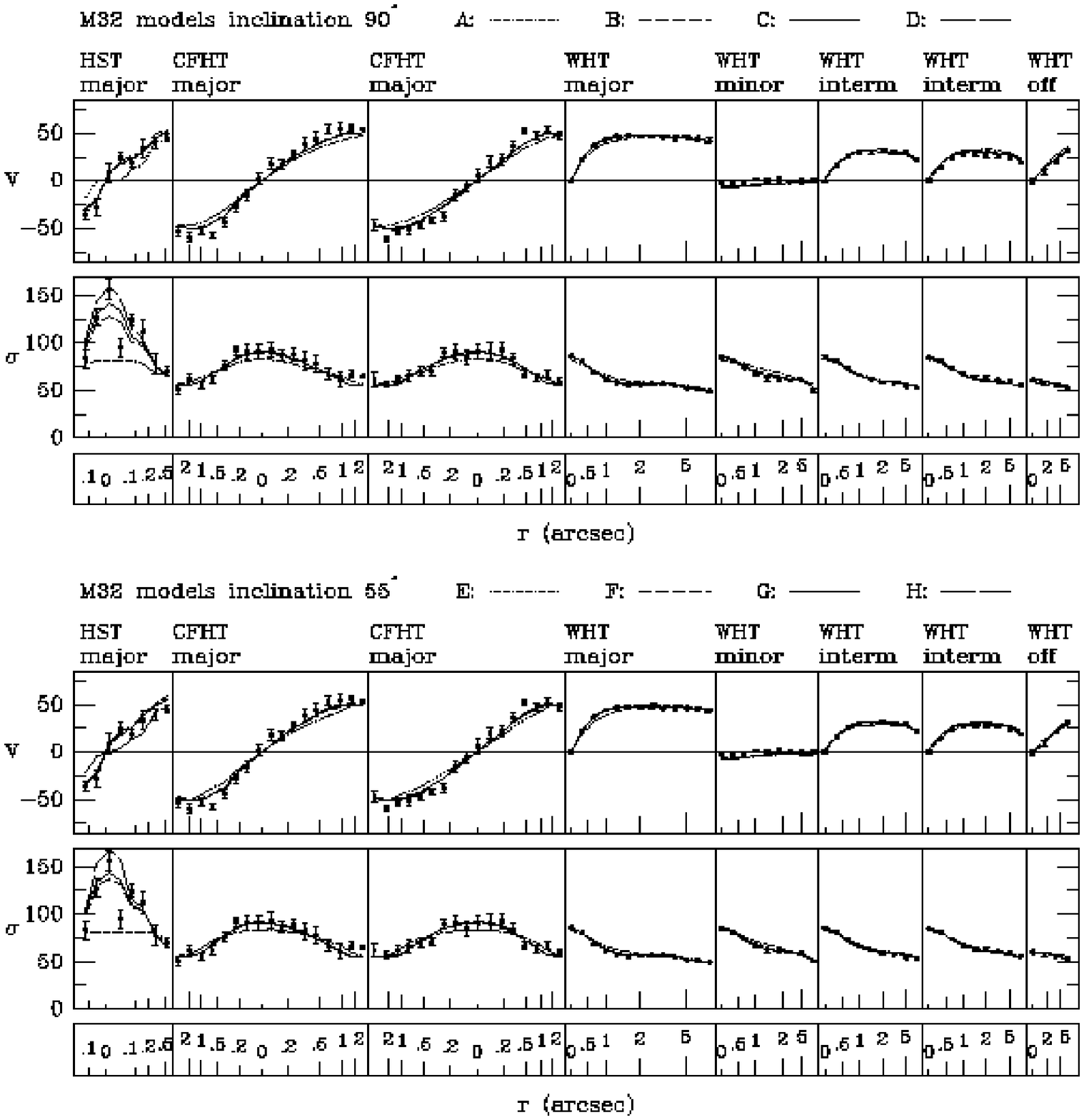}}
\ifsubmode
\vskip3.0truecm
\centerline{Figure~8}\clearpage
\else\figcaption{\figcapeight}\fi
\end{figure}


\begin{figure}
\ifsubmode
\else\figcaption{\figcapnine}\fi
\end{figure}

\begin{figure}
\figurenum{9}
\epsfxsize=12.0truecm
\centerline{\epsfbox{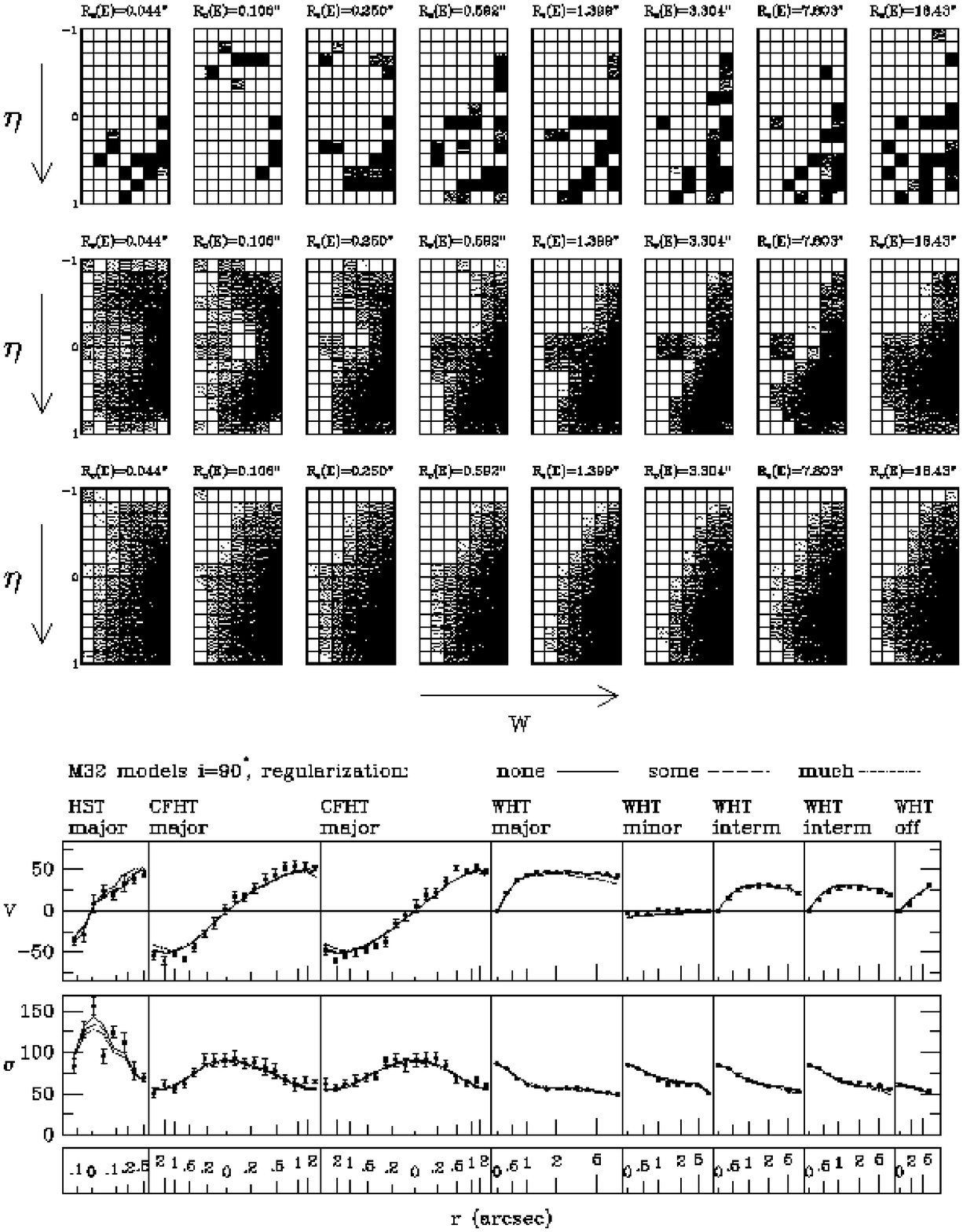}}
\ifsubmode
\vskip3.0truecm
\centerline{Figure~9}\clearpage
\else\figcaption{\figcapnineB}\fi
\end{figure}


\begin{figure}
\centerline{\epsfbox{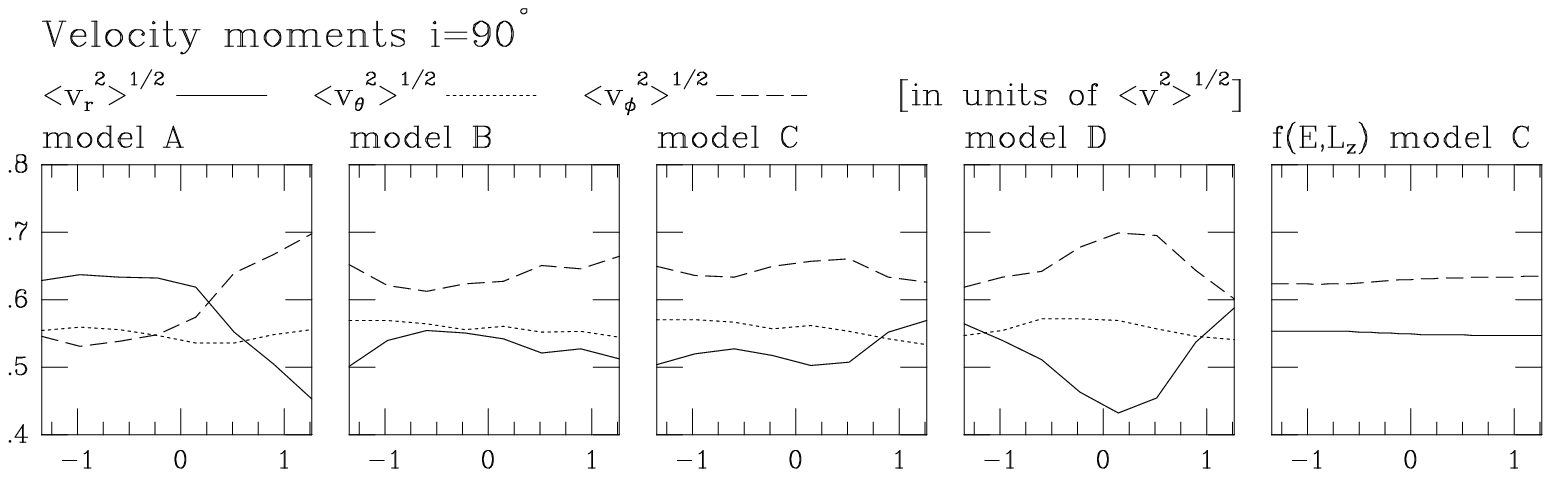}}
\bigskip
\centerline{\epsfbox{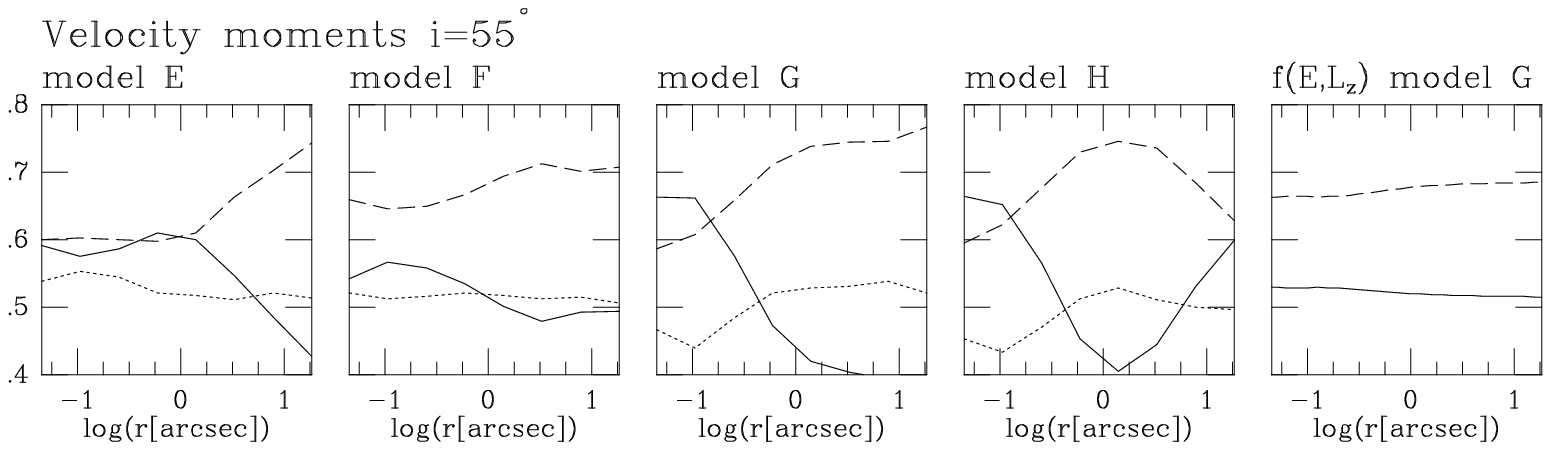}}
\ifsubmode
\vskip3.0truecm
\centerline{Figure~10}\clearpage
\else\figcaption{\figcapten}\fi
\end{figure}


\begin{figure}
\centerline{\epsfbox{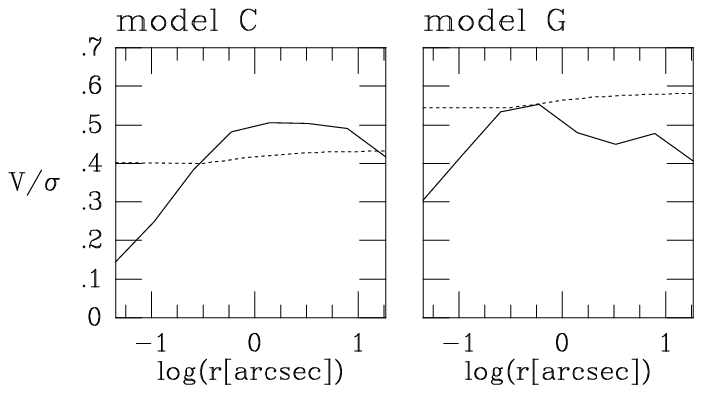}}
\ifsubmode
\vskip3.0truecm
\centerline{Figure~11}\clearpage
\else\figcaption{\figcapeleven}\fi
\end{figure}


\begin{figure}
\centerline{\epsfbox{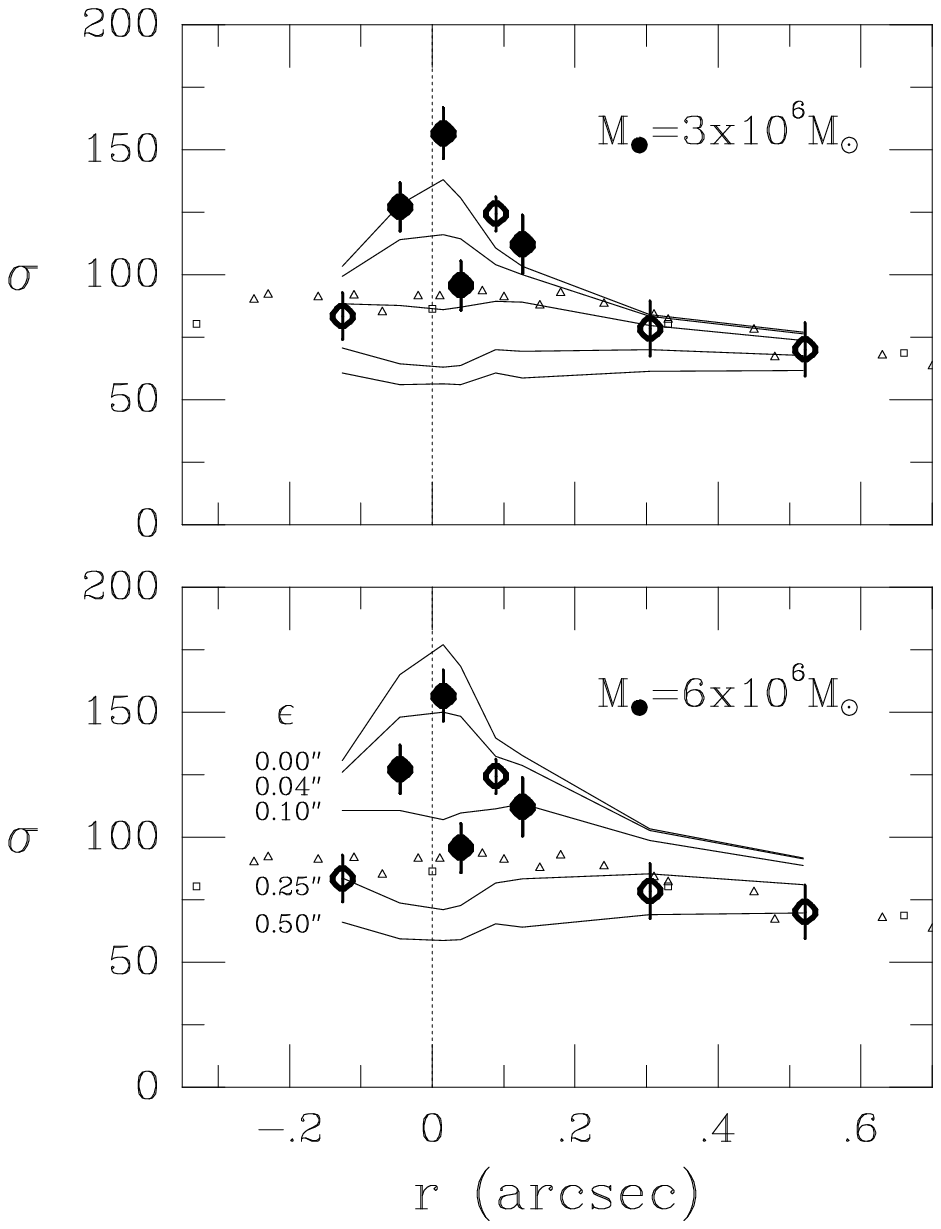}} 
\ifsubmode
\vskip3.0truecm
\centerline{Figure~12}\clearpage
\else\figcaption{\figcaptwelve}\fi
\end{figure}


\begin{figure}
\noindent\begin{minipage}[b]{8.0truecm}
\epsfxsize=8.0truecm
\epsfbox{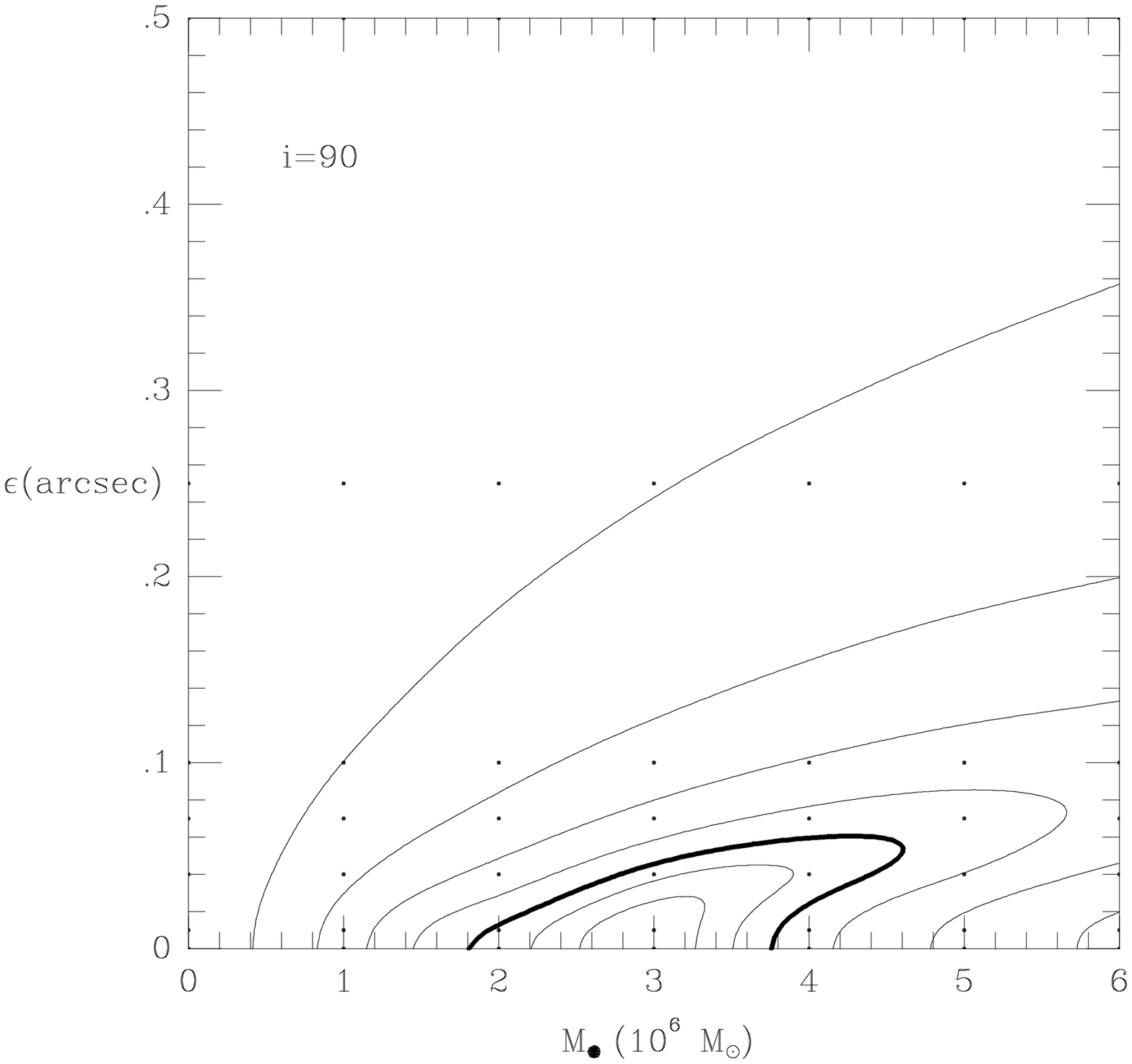}
\end{minipage}
\hfill
\noindent\begin{minipage}[b]{8.0truecm}
\epsfxsize=8.0truecm
\epsfbox{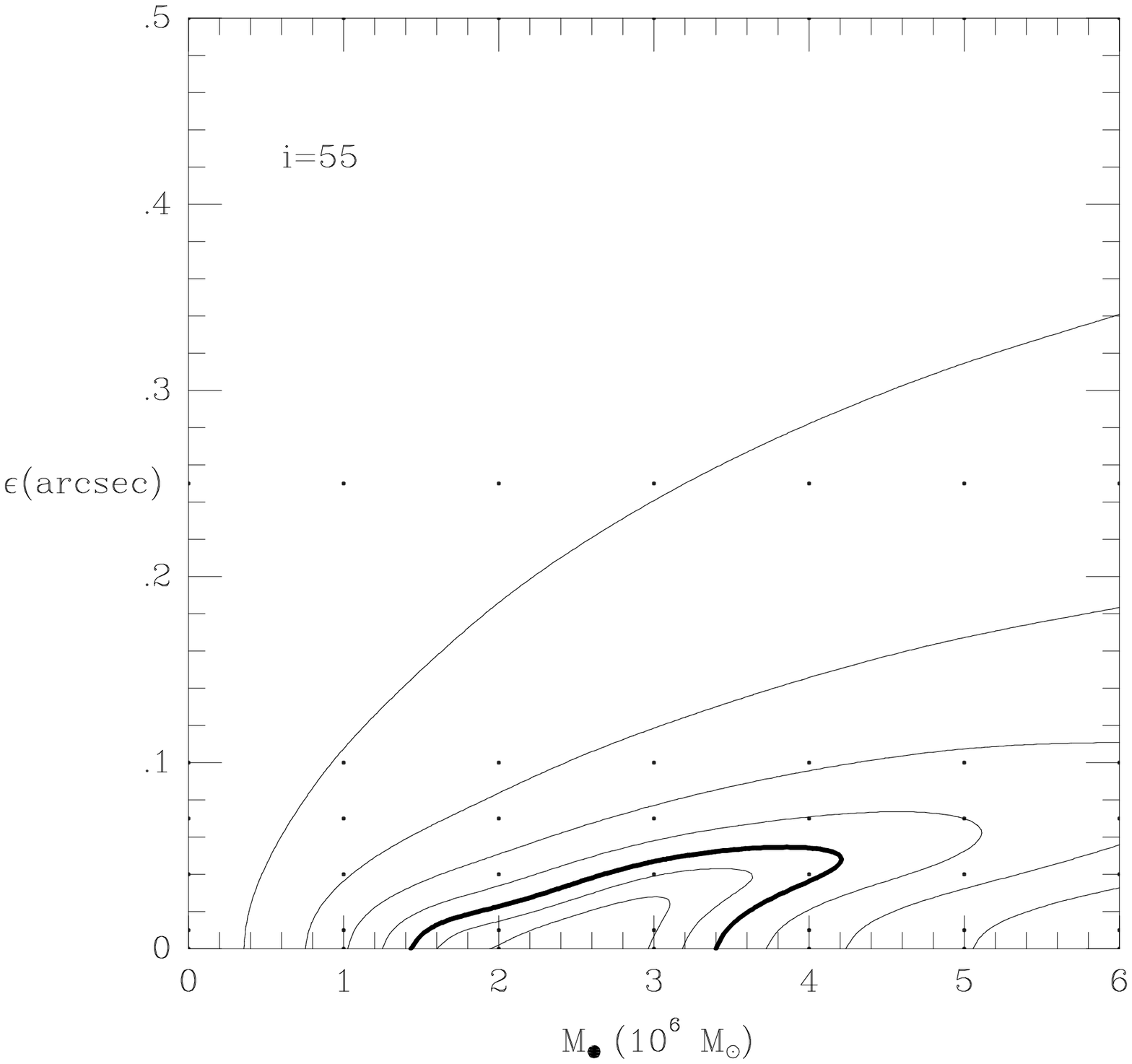}
\end{minipage}
\ifsubmode
\vskip3.0truecm
\centerline{Figure~13}\clearpage
\else\figcaption{\figcapthirteen}\fi
\end{figure}


\begin{figure}
\epsfxsize=16.0truecm
\centerline{\epsfbox{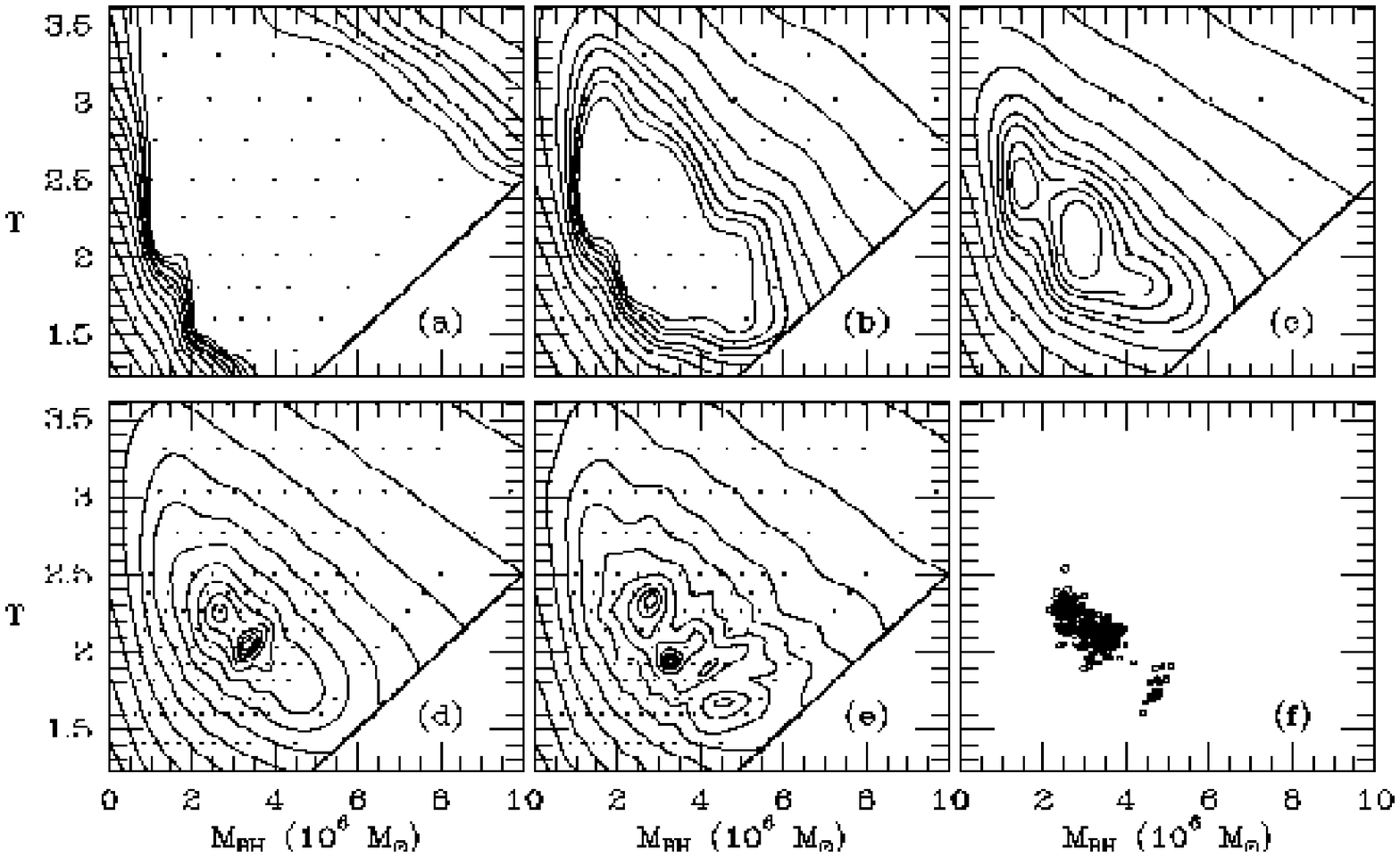}}
\ifsubmode
\vskip3.0truecm
\centerline{Figure~14}\clearpage
\else\figcaption{\figcapfourteen}\fi
\end{figure}


\begin{figure}
\centerline{\epsfbox{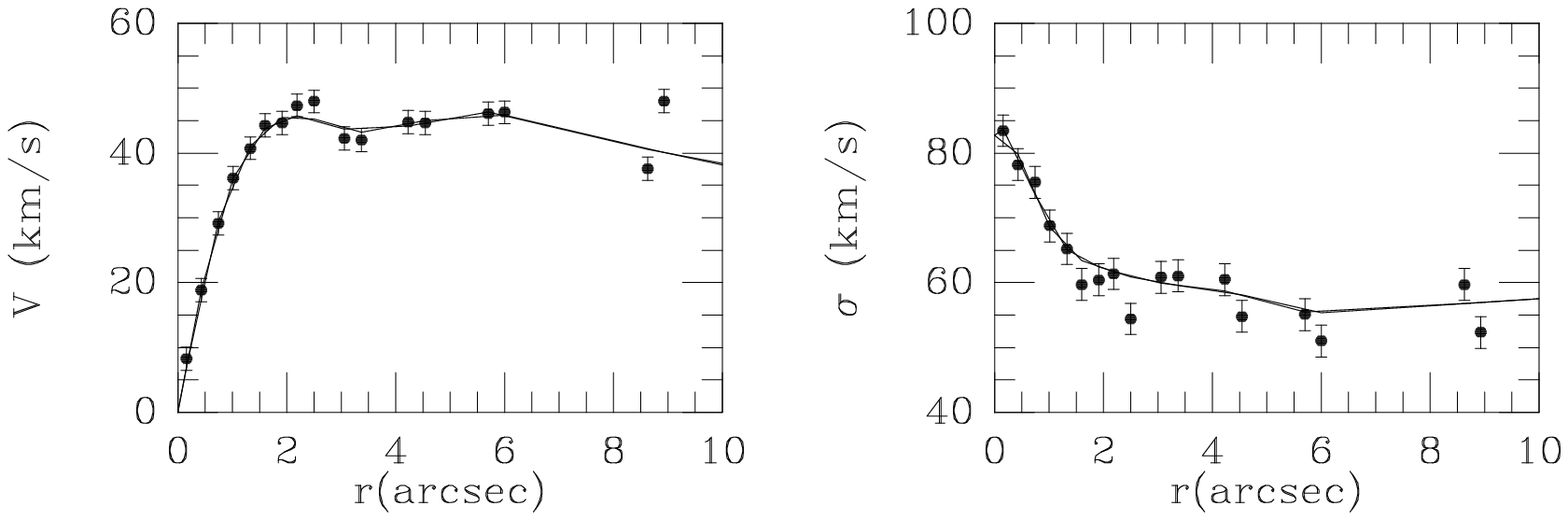}} 
\ifsubmode
\vskip3.0truecm
\centerline{Figure~15}\clearpage
\else\figcaption{\figcapfifteen}\fi
\end{figure}


\fi

\end{document}